\documentclass[
preprint,
 amsmath,amssymb,
 aps,
prb,
longbibliography
]{revtex4-1}

\usepackage{graphicx}
\usepackage{dcolumn}
\usepackage{bm}
\usepackage{color}
\usepackage{braket}
\usepackage{multirow}
\usepackage[colorlinks,linkcolor=blue, urlcolor=blue,citecolor=blue,bookmarks=fales]{hyperref}
\usepackage{lineno}

\begin{document}

\title{Superconductivity near the saddle point in the two-dimensional Rashba system Si(111)-$\sqrt{3}\times\sqrt{3}$-(Tl,Pb)}

\author{T. Machida$^{1,2,\ast}$, Y. Yoshimura$^{3}$, T. Nakamura$^{4}$, Y. Kohsaka$^{1}$, T. Hanaguri$^{1}$, C.-R. Hsing$^{5}$, C.-M. Wei$^{5}$, Y. Hasegawa$^{6}$, S. Hasegawa$^{7}$, and A. Takayama$^{3}$}

\affiliation{$^{1}$RIKEN Center for Emergent Matter Science, Wako 351-0198, Japan}
\affiliation{$^{2}$Precursory Research for Embryonic Science and Technology (PRESTO), Japan Science and Technology Agency (JST), Tokyo 102-0076, Japan}
\affiliation{$^{3}$Department of Physics and Applied Physics, Waseda University, Tokyo 169-8555, Japan}
\affiliation{$^{4}$Center for GaN Characterization and Analysis, National Institute for Materials Science, Tsukuba 305-0047, Japan}
\affiliation{$^{5}$Institute of Atomic and Molecular Sciences, Academia Sinica, P.O. Box 23-166, Taipei, Taiwan}
\affiliation{$^{6}$The Institute for Solid State Physics, The University of Tokyo, Kashiwa 277-8581, Japan}
\affiliation{$^{7}$Department of Physics, University of Tokyo, Tokyo 113-0033, Japan}

\begin{abstract}
Two-dimensional Rashba superconductor Si(111)-$\sqrt{3}\times\sqrt{3}$-(Tl, Pb) is a candidate platform of mixed spin-singlet and -triplet superconductivity.
A recent scanning tunneling microscope (STM) experiment revealed a pseudogap at the vortex core, suggesting the finite triplet component [T. Nakamura {\it et al.},  Phys. Rev. B {\bf 98}, 134505 (2018)].
Detailed spectroscopic information of the superconducting gap and the low-energy band structure is necessary to establish the putative triplet superconductivity. 
Here, we performed high-energy-resolution spectroscopic imaging experiments on Si(111)-$\sqrt{3}\times\sqrt{3}$-(Tl, Pb) using an ultra-low temperature STM.
We found that various spectroscopic features, including the vortex-core spectrum, are consistent with spin-singlet $s$-wave superconductivity, having no sign of the triplet component.
The apparent contradiction with the previous STM result suggests that the nature of superconductivity changes within the same system.
From the analysis of the quasiparticle interference patterns, we found that the Fermi energy is in the close vicinity of the saddle point  near the $\overline{\rm{M}}$ point.
We speculate that the nature of superconductivity varies depending on the saddle-point energy with respect to the Fermi energy, which is sample-dependent due to different band filling.
\end{abstract}

\maketitle

\section{Introduction}
Superconductivity in the systems without inversion symmetry possesses mixed even- and odd-parity superconducting wave functions~\cite{Gorkov_PRL_2001,Sato_PRB_2009,Mizushima_PRB_2014,Yao_PRB_2015,Lu_PRB_2018,Trott_PRR_2020}.
If the odd-parity, namely spin-triplet, component becomes dominant, a topologically nontrivial superconducting state may be realized~\cite{Sato_PRB_2009,Mizushima_PRB_2014,Yao_PRB_2015}. Therefore, vigorous efforts have been made to induce superconductivity in the system with broken inversion symmetry, as exemplified by Rashba semiconductor nanowires on superconductors \cite{Mourik_Science_2012,Zhang_Nature_2018,Lutchyn_NRM_2018}, Rashba-material-superconductor heterostructures~\cite{Manna_PNAS_2020, Kezilebieke_Nature_2020}, and superconductors with topological surface states~\cite{Zhang_Science_2018,Wang_Science_2018,Machida_NM_2019,Iwaya_NC_2017,Guan_SAvd_2018}.
Another intriguing example is an atomically thin 2-dimensional (2D) superconductor composed of heavy elements grown on an insulating or semiconducting substrate~\cite{Zhang_NP_2010,Yoshizawa_PRL_2014,Yoshizawa_NC_2021,Matetskiy_PRL_2015}.
In such a system, the combination of the strong spin-orbit coupling and the broken inversion symmetry perpendicular to the interface leads to a Rashba-type spin splitting in the band structure.
Among such 2D superconductors, Si(111)-$\sqrt{3}\times\sqrt{3}$-(Tl, Pb) has attracted attention~\cite{Gruznev_SRep_2014,Matetskiy_PRL_2015,Nakamura_PRB_2018}.
Recent scanning tunneling microscopy (STM) experiments have revealed signatures that suggest spin-triplet superconductivity, such as the vortex density lower than that expected from the applied magnetic field and the pseudogap in the spectrum at the vortex core~\cite{Nakamura_PRB_2018}.
While these observations may be consistent with the theoretical prediction of the doubly quantized vortex core in a spin-triplet chiral $p$-wave superconductor~\cite{Sauls_NJP_2009}, detailed structures of the superconducting gap and the vortex core are yet to be clarified.
In addition, it is important to investigate the band and spin structures near the Fermi energy ($E_{\rm F}$), to reveal the features that may be responsible for the putative spin-triplet pairing.

To this end, we performed spectroscopic-imaging STM on Si(111)-$\sqrt{3}\times\sqrt{3}$-(Tl, Pb) using a home-built dilution-refrigerator STM that can achieve an effective electron temperature below 90~mK corresponding energy resolution of $\sim 20$~$\mu$eV~\cite{Machida_RSI_2018}.
Surprisingly, the superconducting gap spectrum and its temperature dependence, the vortex density, and the spectrum at the vortex core are all qualitatively distinct from the results of previous STM measurements~\cite{Nakamura_PRB_2018} and are consistent with those expected for spin-singlet $s$-wave superconductivity.
To clarify the origin of the discrepancy, we investigated the detailed band structure near the $E_{\rm F}$ by a combination of quasiparticle interference (QPI) imaging and numerical simulations.
We found that the Fermi level lies in the close vicinity of the saddle point of the spin-split band near the $\overline{\rm{M}}$ point.
These QPI results suggest that even a small difference in the saddle point energy with respect to $E_{\rm F}$ alters topology and spin structures of the Fermi surface through the Lifshitz transition.
We speculate that the qualitatively distinct superconducting properties observed in Si(111)-$\sqrt{3}\times\sqrt{3}$-(Tl, Pb) are related to this Lifshitz transition.

\section{Methods}
Si(111)-$\sqrt{3}\times\sqrt{3}$-(Tl, Pb) samples were grown in an ultra-high vacuum (UHV) chamber attached to the ultra-low temperature STM system. (base pressure $\sim 1\times10^{-10}$~Torr).
The Si substrate was cut from an As-doped n-type wafer (room temperature resistivity of a few m$\rm{\Omega}$cm) taken from the same lot used in Ref. \onlinecite{Nakamura_PRB_2018}.
We first prepared a clean Si(111)-7$\times$7 reconstructed surface by flashing the wafer at 1135~$^{\circ}$C several times.
After the final flashing, we slowly cooled the wafer from 850~$^{\circ}$C to 600~$^{\circ}$C at $\sim -0.5$~$^{\circ}$C/sec.
Subsequently, monolayer Tl (purity: 99.99\%) was deposited from a Knudsen cell with keeping the substrate temperature at 300~$^{\circ}$C. 
One-third of monolayer Pb (purity: 99.9999\%) was then deposited at room temperature using a water-cooled Knudsen cell.
We carefully optimized the amount of Tl and Pb so that the area of the $\sqrt{3}\times\sqrt{3}$-(Tl, Pb) structure was maximized in STM topographic images.
The samples were transferred to the STM head immediately after the growth.

We performed STM experiments using the dilution-refrigerator STM mentioned above\cite{Machida_RSI_2018}.
The STM tip was an electrochemically-etched tungsten wire cleaned by electron-beam heating followed by field evaporation using a field-ion microscope. 
All the spectroscopic experiments were done using the standard lock-in method with a modulation frequency of 614.7~Hz. 
The bias voltage was applied to the sample.

\section{Results and Discussion}
Figure~1(a) shows a typical topographic image in a $1~\mu\rm{m}~\times~1~\mu\rm{m}$ field of view, showing the step-terrace structures of the Si substrate.
On each terrace, $\sqrt{3}\times\sqrt{3}$-(Tl, Pb) monolayer is formed [Fig.~1(b)].
An atomic-resolution topographic image shown in Fig.~1(c) is characterized by a hexagonal structure with a period of $\sqrt{3}a_{0}$ ($a_{0}=3.84$~\AA\ is the lattice constant of the Si(111) surface without the reconstruction).
The observed topographic features agree well with the previous STM results~\cite{Nakamura_PRB_2018} and are consistent with the structure model shown in Fig.~1(c) inset.

Figure~1(d) depicts a tunneling spectrum over a wide energy range.
We also show the spectrum taken in Ref.~\onlinecite{Nakamura_PRB_2018} for comparison.
Both spectra are characterized by a sharp peak around -400~meV, which may be related to the Si(111) substrate, and a hump around +150~meV.
Although the peak and hump are well reproduced, these features in the present sample shift downward compared to those in Ref.~\onlinecite{Nakamura_PRB_2018} [Fig.~1(e) and 1(f)].
We note that the samples used in this work and in Ref. ~\onlinecite{Nakamura_PRB_2018} were prepared in the different UHV systems even though we used the Si substrate taken from the same lot. 
Therefore subtle differences in the effective doping level and the Tl/Pb ratio may result, causing the energy shift in the spectrum.
While its exact origin is elusive, the observed downward shift suggests that the energy bands of the present sample are filled with more electrons than in Ref ~\onlinecite{Nakamura_PRB_2018}.

Next, we examined the superconducting properties.
Figure 2(a) shows superconducting-gap spectra at different temperatures.
The spectrum at the lowest temperature exhibits the fully gapped behavior, indicating that the superconducting gap is isotropic. 
The isotropic gap is in contrast to the anisotropic gap observed in Ref.~\onlinecite{Nakamura_PRB_2018}.
We note that this discrepancy is not due to the inhomogeneity of the sample because the spectrum is uniform over $\sim 150$~nm, as shown in Fig.~2(b).
The observed fully-gapped spectrum can be fitted well by the Dynes function that includes the thermal broadening and the bias modulation of the lock-in measurement.
The parameters obtained by the fitting are summarized in the Table I.
From the fitting results at different temperatures, we determined the temperature $T$ dependence of the superconducting gap $\Delta(T)$, as shown in Fig.~2(c).
The obtained $\Delta(T)$ well follows the behavior expected from the conventional Bardeen-Cooper-Schrieffer theory.
We estimated the gap ratio $2\Delta(0)/k_{\rm B}T_{\rm c}$ to be 4.4.
Here, $k_{\rm B}$ is the Boltzmann constant, and $T_{\rm c}$ is the superconducting transition temperature defined at the gap-closing point.
The gap ratio indicates the strong-coupling nature of superconductivity but is about a half of $2\Delta(0)/k_{\rm B}T_{\rm c} \sim 8.6$ estimated in Ref.~\onlinecite{Nakamura_PRB_2018}.

Different superconducting properties between present and previous results~\cite{Nakamura_PRB_2018} are also seen in the vortex state in a magnetic field.
Figure 3(a) represents a vortex image in a magnetic field of 0.1~T applied perpendicular to the surface.
This is a current map $I(\bm{r},~E=\rm{+300~\mu eV})$ corresponding to the energy ($E$) integrated local density of states from $E_{\rm F}$ to +300$~\mu$eV. 
The current value at each pixel was acquired by setting the sample bias voltage $V_{\rm{s}}$ to +300~$\mu$V after turning off the feedback-loop for scanning.
There are 48 vortices (bright regions) in this image, consistent with the number expected from the external magnetic field and the conventional singly quantized vortex.
Vortices distribute inhomogeneously due to the vortex pinning by defects such as the clusters of excess Tl and step edges.
We then acquired a tunneling spectrum at a vortex core marked by a circle in Fig.~3(a).
As shown in Fig.~3(b), a pseudogap at the vortex core~\cite{Nakamura_PRB_2018} was not detected.
Instead, a broad peak appears near zero energy.
With increasing distance from the vortex center, this peak gets broadened and evolves into the spectrum with a clear superconducting gap, as shown in Fig.~3(b) and (c). 
This is the conventional behavior expected for Caroli-de Gennes-Matricon bound states in the vortex core~\cite{CdGM_PhysLett_1964}.

Summarizing so far, all of the observed spectroscopic features in the present sample are consistent with the behaviors expected in a spin-singlet isotropic $s$-wave superconductor, even though the possibility of particular fully-gapped spin-triplet superconductivity cannot be completely ruled out. The important point is that the superconducting properties of the present samples are qualitatively different from the previous results reported in Ref. \onlinecite{Nakamura_PRB_2018}.
Such contrasting behaviors suggest that Si(111)-$\sqrt{3}\times\sqrt{3}$-(Tl, Pb) can host different superconducting states in the same system, raising the question of what parameters tune the nature of superconductivity.
A natural candidate would be the band filling as suggested from the shifts in the peak and hump in the spectra [Figs.~1(e) and 1(f)].
For instance, if the topology and spin structures of the Fermi surface vary upon band filling, the nature of superconductivity may also change. 
Here we assume that saddle points near the $\overline{\rm{M}}$ point in Si(111)-$\sqrt{3}\times\sqrt{3}$-(Tl, Pb) may play a role in such a change.

Figures~4(a)-(d) show the band dispersions and constant energy contours near $E_{\rm{F}}$ calculated from a model tight-binding Hamiltonian based on the results of angle-resolved photoemission spectroscopy~\cite{Matetskiy_PRL_2015,Gruznev_SRep_2014} and first-principles calculations~\cite{Gruznev_SRep_2014} (See Appendix B for the details).
At low energies, there are two concentric $\overline{\rm{\Gamma}}$-centered bands, each of which exhibits Rashba spin splitting. 
The outer band reaches the $\overline{\rm{M}}$ point upon increasing energy, causing the saddle point in the band dispersion. 
The topology of the constant energy contour changes from the $\overline{\rm{\Gamma}}$-centered hexagon to the $\overline{\rm{K}}$-centered triangle, exemplifying a platform for the Lifshitz transition.
The Rashba effect makes the spin structure complicated and splits the saddle point into two types at $E_{\rm{c1}}$ and $E_{\rm{c2}}$.
Since $E_{\rm{c1}}$ and $E_{\rm{c2}}$ are located only a few tens of meV above $E_{\rm F}$, even a small change in the band filling may cause the Lifshitz transition that accompanies the changes in the topology and spin structures of the Fermi surface. 
Moreover, the van Hove singularity (VHS) at the saddle point may dominate the low-energy density of states, thereby playing an essential role in the pairing interaction.

The experimental challenge is how to detect the saddle point above $E_{\rm F}$ where the angle resolved photoemission spectroscopy is not available. Here we propose that the quasiparticle interference (QPI) imaging provides a sensitive probe of the saddle point in the band dispersion regardless of its energy location. 
Before considering the QPI patterns in the actual band structure with Rashba spin splitting, we outline the key features of the QPI patterns that signify the saddle point using a single spin-degenerated band model [Fig. 4(e)].
As we will describe later, this simplified model captures the basic features of the QPI patterns even in the case of the Rashba-split band.
If the energy increases toward the saddle point, a pair of arcs in the constant-energy contour [blue arcs in Fig. 4(e)] comes closer and merges at the saddle point. 
If the energy goes beyond the saddle point, another pair of arcs [red arcs in Fig. 4(e)] appears along the orthogonal axis and departs from each other.
We expect quasiparticle scattering processes that connect these arcs, as shown by the red and blue arrows in Fig.~4(e).
As a result, hole-like and electron-like QPI branches that disperse along the orthogonal axes appear, and they converge at the saddle-point energy.
The expected QPI wave vector $\bm{q}$ is inevitably small and unavoidable inhomogeneities etc. in a real sample may make it difficult to detect such QPI signals near $\bm{q}$~=~0.
Therefore, practically, it is easier to search for the "replica" QPI signals near the Bragg point $\overline{\rm{\Gamma'}}$ where the QPI signals shifted by the reciprocal lattice vector $\bm{G}$ is expected [Fig. 4(f) and 4(g)].

We performed QPI-imaging experiments in the normal state at 4.4~K.
Figures~5(a)-(c) represent the Fourier-transformed conductance maps in the same field of view as Fig. 1(b) showing the QPI signals.
The high symmetry points outside of the first Brillouin zone are denoted as $\overline{\rm{\Gamma'}}$, $\overline{\rm{M'}}$, and $\overline{\rm{K'}}$.
Below $E_{\rm F}$, arc-shaped features are seen near the $\overline{\rm{\Gamma'}}$point along the $\overline{\rm{\Gamma'}}\overline{\rm{M}}$ axis [Fig. 5(d)].
These arcs come closer, merge near $E_{\rm F}$, and change to another pair of arcs moving away along the orthogonal  $\overline{\rm{\Gamma'}}\overline{\rm{K'}}$ axis above $E_{\rm F}$ [Fig. 5(e) and 5(f)].
Line-cuts from the Fourier-transformed QPI patterns [Figs.~5(g) and 5(h)] exhibit electron-like and hole-like QPI branches that converge near $E_{\rm F}$.
These observations are exactly what we expect near the saddle point.

Although the simple saddle-point model qualitatively captures the observed QPI patterns, the effect of Rashba splitting should be considered for the quantitative analysis.
We should consider both intra- and inter-branch quasiparticle scattering processes and the spin selection rule for QPI.
Therefore, we performed a QPI simulation with these effects taken into account, based on the $T$-matrix formalism using the tight-binding band structure shown in Fig. 4(a)-(d) (details are given in the Appendix B).

The calculated QPI patterns are shown in Fig.~6(a)-(c).
Because the spin-selection rule suppresses the scattering channels connecting the two states with opposite spin directions, the calculated QPI patterns remain relatively simple despite the Rashba-split multiple constant energy contours.
Pairs of arcs appear along the $\overline{\rm{\Gamma'}}\overline{\rm{M}}$ and  $\overline{\rm{\Gamma'}}\overline{\rm{K'}}$ axes below and above the saddle-point energies, respectively [Figs.~6(d) and 6(f)].
To identify the saddle point energies where the pairs merge, we take line-cuts as shown in Figs.~6(g) and 6(h).
There are multiple hole-like and electron-like branches, but the merging of the hole-like branch along $\overline{\rm{\Gamma'}}\overline{\rm{M}}$ and the electron-like branch along  $\overline{\rm{\Gamma'}}\overline{\rm{K'}}$ occurs at $E_{\rm{c1}}$ as highlighted by the green dashed lines in Figs.~6(g) and 6(h).
Therefore, we conclude that the experimentally observed QPI branches in Fig.~5(g) and 5(h) are the ones merging at $E_{\rm{c1}}$, which almost coincides with $E_{\rm F}$ in the present sample.
Judging from the energy shifts in the tunneling spectra [Figs.~1(e) and 1(f)], we estimate that the $E_{\rm F}$ of the sample investigated in Ref. \onlinecite{Nakamura_PRB_2018} is $\sim 15$~meV below $E_{\rm{c1}}$.
These observations mean that the topology and spin structures of the Fermi surface near the $\overline{\rm{M}}$ point are indeed different between the samples investigated in this work and in Ref.~\onlinecite{Nakamura_PRB_2018}, corroborating the conjecture that the saddle point plays some roles in the pairing interaction.

How does the close vicinity of the saddle point to the $E_{\rm F}$ result in the difference of superconducting properties? 
Recently, the role of VHS at the saddle point has been argued theoretically in the 2D-Rashba superconductors~\cite{Yao_PRB_2015,Greco_PRL_2018,Lu_PRB_2018,Ghadimi_PRB_2019,Qin_NP_2019,Trott_PRR_2020,Nogaki_PRB_2020}.
The VHS can be classified into two types according to the location of the saddle point in momentum $\bm{k}$ space.
The VHS is called type I if the saddle point is located at the time-reversal invariant momenta ($\bm{k} \equiv -\bm{k} + \bm{G}$), whereas it is type II otherwise.
Functional renormalization group analyses suggest that spin-triplet superconductivity could be stabilized when the type II VHS plays a role~\cite{Yao_PRB_2015,Trott_PRR_2020,Qin_NP_2019}.
On the contrary, type I VHS does not contribute to inducing spin-triplet superconductivity, because the time-reversal symmetry demands the superconducting gap to be even parity $\Delta(\bm{k}) = \Delta(-\bm{k})$~\cite{Yao_PRB_2015}.
Thus, depending on the type of the VHS near $E_{\rm F}$, different parity may be favored.
In the case of Si(111)-$\sqrt{3}\times\sqrt{3}$-(Tl,Pb), the $E_{\rm F}$ of the present sample is at the VHS at $E_{\rm{c1}}$.
However, this VHS may not be responsible for the observed spin-singlet behaviors because it is of type II.
The putative spin-triplet superconductivity in the sample of Ref.~\onlinecite{Nakamura_PRB_2018} may also not be related to the VHS, because type II VHS does not exist $\sim 15$~meV below $E_{\rm{c1}}$ where $E_{\rm F}$ is located.
Therefore, the VHS-induced pairing mechanism alone cannot explain the observed change in the dominant character of superconductivity between triplet and singlet.
More detailed theoretical analysis that takes into account realistic band structure etc. is indispensable to understand the nature of superconductivity in Si(111)-$\sqrt{3}\times\sqrt{3}$-(Tl,Pb).

\section{Conclusion}
We have performed spectroscopic-imaging STM measurements on Si(111)-$\sqrt{3}\times\sqrt{3}$-(Tl, Pb) to investigate its superconducting properties and the low-energy band structure.
We found that the superconducting-gap spectrum and its temperature dependence, the density of vortices, and the spectrum at the vortex core, are consistent with those of the spin-singlet conventional $s$-wave superconductor.
This is in contrast to the previous STM results on the same system where the signatures of spin-triplet $p$-wave superconductivity were observed~\cite{Nakamura_PRB_2018}.
We speculate that the observed different superconducting properties are related to the band-filling difference, which manifests in a slight energy shift of the tunneling spectrum.
QPI measurements along with the theoretical simulations show that the Fermi level of the present sample lies very close to the saddle point of the spin-split bands near the $\overline{\rm{M}}$ point.
These results evidence that even a tiny change in the band filling can induce the qualitative changes of the Fermi-surface topology and the spin textures around $E_{\rm{F}}$.
Even though the microscopic mechanism that gives rise to different superconducting states in the same system is yet to be clarified, our results suggest that Si(111)-$\sqrt{3}\times\sqrt{3}$-(Tl, Pb) is a unique platform that enables us to control the nature of superconductivity through a slight change of the band filling.

\section*{Acknowledgments}
This work was supported by CREST Project No. JPMJCR16F2, PRESTO project No. JPMJPR19L8 from JST 
and Grants-in-Aid for Scientific Research (KAKENHI) (numbers 19H01843 and 20H00342).
We appreciate T. Sato and K. Sugawara for providing us with the evaporation source of Pb.
We also thank Y. Tsutusumi for valuable discussions.

\appendix
\section{Parameters obtained by fitting the tunneling spectrum in Fig. 2(a) with Dynes function}
To obtain the temperature dependence of the superconducting gap size ($\Delta$), we fit the tunneling spectra at several temperatures in Fig. 2(a) by using Dynes function,
\begin{equation}
    \rho(E) = \mathrm{Re}\left(\frac{E-i\Gamma}{(\sqrt{(E-i\Gamma)^2-\Delta^2}}\right),
\end{equation}
where $\rho$ and $\Gamma$ are density-of-states and the phenological energy broadening, respectively.
Table I summarizes the obtained parameters.
\begin{table*}[htb]
    \caption{Parameters determined by fitting the tunneling spectra at several temperatures under 0~T.}
    {\tabcolsep = 1.5mm
      \begin{tabular}{ccc}
    \hline
    \hline
      Temperature &  $\Delta$ & $\Gamma$ \\
      \hline
      0.09~K & 644.1 $\pm$ 0.7 $\mu$eV & 17.6 $\pm$ 0.7 $\mu$eV \\
      0.25~K & 641.4 $\pm$ 0.7 $\mu$eV & 7.89 $\pm$ 0.7 $\mu$eV \\
      0.5~K & 638.4 $\pm$ 0.1 $\mu$eV & 1.0 $\pm$ 0.8 $\mu$eV \\
      0.77~K & 629.2 $\pm$ 0.8 $\mu$eV & 1.2 $\pm$ 0.9 $\mu$eV \\
      1.44~K & 618.3 $\pm$ 1.2 $\mu$eV & 1.1 $\pm$ 1.2 $\mu$eV \\
      2.0~K & 589.2 $\pm$ 1.0 $\mu$eV & 12.5 $\pm$ 1.0 $\mu$eV \\
      2.3~K & 553.9 $\pm$ 1.1 $\mu$eV & 18.1 $\pm$ 1.2 $\mu$eV \\
      2.6~K & 497.1 $\pm$ 1.7 $\mu$eV & 21.1 $\pm$ 2.0 $\mu$eV \\
      2.85~K & 430.9 $\pm$ 2.2 $\mu$eV & 22.2 $\pm$ 3.2 $\mu$eV \\
      3.0~K & 376.6 $\pm$ 2.5 $\mu$eV & 21.6 $\pm$ 4.4 $\mu$eV \\
      3.15~K & 311.6 $\pm$ 3.0 $\mu$eV & 23.4 $\pm$ 6.6 $\mu$eV \\
      3.3~K & 231.9 $\pm$ 4.2 $\mu$eV & 53.0 $\pm$ 12.9 $\mu$eV \\
    \hline
    \hline
      \end{tabular}
    }
\end{table*}

\section{Tight-binding Hamiltonian and QPI simulations}
For the QPI simulations shown in the main text, we use a tight-binding Hamiltonian to describe electronic states near $E_{\rm F}$. 
According to the previous density-functional-theory calculation, the contributions from the Tl and Pb atoms dominate the electronic states around $E_{\rm F}$~\cite{Gruznev_SRep_2014}.
Therefore, we consider the electrons traveling in the Tl$_{3}$Pb overlayer.
We adopt the same kinetic term of the Hamiltonian for Pb$_{3}$Bi on Ge(111)~\cite{Sun_PRB_2021} that possesses a similar crystal structure.

\begin{eqnarray}
	\label{eq:Hami_0}
	\hat{H}_{0} &=& E_0 + \sum_{ij}\left(\mu_{1}\delta_{ij}-t_{1}\delta_{\braket{ij}}-t'_{1}\delta_{\braket{\braket{ij}}}\right)a^{\dag}_{i}a_{j} \nonumber \\ 
    &-&\sum_{il}\left(t_{2}\delta_{\braket{il}}+t'_{2}\delta_{\braket{\braket{il}}}\right)a^{\dag}_{i}b_{l} + \mu_{2}\sum_{l} b^{\dag}_{l}b_{l}.
\end{eqnarray}
Here, $a_{i} =(a_{i\uparrow}, a_{i\downarrow})^{T}$ and $b_{l} =(b_{l\uparrow},b_{l\downarrow})^{T}$ represent the annihilation operators on Tl and Pb atomic sites $i$ and $l$, respectively, $E_{0}$ is an energy offset, $\mu_{1,2}$ are the on-site energy on Tl and Pb sites, $t_{1,2}$ and $t'_{1,2}$ are the nearest neighbor (NN) and next-nearest neighbor (NNN) hopping energies (Fig. 7).
The subscript 1(2) represents the Tl-Tl (Pb-Tl) hopping, $\braket{ij}$ and $\braket{\braket{ij}}$ indicate that $i$ and $j$ are NN and NNN sites, respectively.
Considering that the electrons in the Tl$_{3}$Pb overlayer experience the substrate-induced electric field (potential), we write the spin-orbit coupling term as
\begin{eqnarray}
	\label{eq:Hami_SOC}
	\hat{H}_{\rm{SOC}} &=& i\sum_{ij}\left(\lambda_{1}\delta_{\braket{ij}}+ \lambda'_{1}\delta_{\braket{\braket{ij}}}\right)\left(\bm{E}_{ij}\times\bm{r}_{ij}\right)\cdot\bm{\sigma}a^{\dag}_{i}a_{j} \nonumber \\
    &+& i\sum_{il}\left(\lambda_{2}\delta_{\braket{il}}+ \lambda'_{2}\delta_{\braket{\braket{il}}}\right)\left(\bm{E}_{il}\times\bm{r}_{il}\right)\cdot\bm{\sigma}a^{\dag}_{i}b_{l}, \nonumber \\
    &&
\end{eqnarray}
where $\lambda_{1,2}$ and $\lambda'_{1,2}$ are the magnitude of the spin-orbit coupling for the NN and NNN hopping processes.
The vectors $\bm{E}_{ij}$, $\bm{r}_{ij}$, and $\bm{\sigma}$ denote an electric field that the electrons experience during the hopping from $i$ to $j$ sites, a displacement vector from $i$ to $j$ sites, and the Pauli matrix vector, respectively.

We consider not only the out-of-plane but also the in-plane components of the electric fields for all hopping processes.
Since the in-plane component parallel to the corresponding displacement vector does not yield any effect, we consider the in-plane component perpendicular to the displacement vector in each hopping process.
Eventually, we have five parameters associated with the electric fields, including spatially uniform out-of-plane component ($\epsilon_{z}$) and four in-plane components for each hopping process ($\epsilon_{1,2}$ and $\epsilon'_{1,2}$) being necessary to reproduce the canting of the spins along the out-of-plane direction as shown in Fig.~4, where the absence (presence) of the prime indicates the NN (NNN) hopping and the subscript 1(2) denotes the Tl-Tl (Pb-Tl) hopping.

For simplicity, we set the NN Tl-Tl distance ($a$) and the spatially uniform out-of-plane electric field $\epsilon_{z}$ to be unity.
To obtain the eigenvalues $E_{\bm{k}}$ and eigenvectors $\ket{\psi(\bm{k})}$, we diagonalize the single-particle Hamiltonian in momentum space $H(\bm{k})$,
which is obtained by performing Fourier transformation of $\hat{H} = \hat{H}_{0} + \hat{H}_{\rm{SOC}}$.
We determine the parameters in this Hamiltonian so that the resulting band structure in the vicinity of $E_{\rm F}$ and the spin texture on the Fermi surface agree well with those of the density-functional-theory calculation~\cite{Gruznev_SRep_2014}.
Table II lists the parameters used in this work.

\begin{table*}[htb]
    \caption{Tight-binding parameters used in the QPI simulation.
    For $E_0$, $\mu_{1,2)}$, $t^{(')}_{1,2}$, $\lambda^{(')}_{1,2}$, the units is eV.
    For $\epsilon^{(')}_{1,2}$, the unit is $\epsilon_{z}$ which is set to be unity.}
    {\tabcolsep = 1.5mm
      \begin{tabular}{cccccccc}
    \hline
    \hline
      $E_0$ &  $\mu_{1}$ & $\mu_{2}$ & $t_{1}$ & $t_{2}$ & $t'_{1}$ & $t'_{2}$ & -  \\
      -0.281 & 0.870 & 0.324 & -0.814 & 0.430 & 0.052 & 0.370 & -  \\
      \hline
      $\lambda_{1}$ & $\lambda_{2}$ & $\lambda'_{1}$ & $\lambda'_{2}$ & $\epsilon_{1}$ & $\epsilon_{2}$ & $\epsilon'_{1}$ & $\epsilon'_{2}$ \\
      0.023 & 0.035 & -0.001 & -0.034 & 0.521 & 0.000 & 0.100 & 1.206 \\
    \hline
    \hline
      \end{tabular}
    }
\end{table*}
    
For the QPI simulation, we employ the standard $T$-matrix formalism [See equations (2)-(4) in Ref.~\onlinecite{Kohsaka_PRB_2017}].
We assume a momentum-independent non-magnetic scalar scattering potential,
\begin{equation}
    \label{eq:Vmatrix}
    \hat{V}_{\bm{k},\bm{k}'}  = V_{0}\hat{I},
\end{equation}
where $\hat{I}$ denotes the identity matrix and we set $V_0~=~0.01~eV$.
We consider the so-called set-point effect~\cite{Kohsaka_Science_2007} and the lock-in broadening (0.88~mV$_\textrm{rms}$) to compare the simulated results with the experimental data.
Details of the QPI simulation are described in Ref.~\onlinecite{Kohsaka_PRB_2017}.

\bibliography{PbTl_on_Si_Refs}

\begin{thebibliography}{33}%
\makeatletter
\providecommand \@ifxundefined [1]{%
 \@ifx{#1\undefined}
}%
\providecommand \@ifnum [1]{%
 \ifnum #1\expandafter \@firstoftwo
 \else \expandafter \@secondoftwo
 \fi
}%
\providecommand \@ifx [1]{%
 \ifx #1\expandafter \@firstoftwo
 \else \expandafter \@secondoftwo
 \fi
}%
\providecommand \natexlab [1]{#1}%
\providecommand \enquote  [1]{``#1''}%
\providecommand \bibnamefont  [1]{#1}%
\providecommand \bibfnamefont [1]{#1}%
\providecommand \citenamefont [1]{#1}%
\providecommand \href@noop [0]{\@secondoftwo}%
\providecommand \href [0]{\begingroup \@sanitize@url \@href}%
\providecommand \@href[1]{\@@startlink{#1}\@@href}%
\providecommand \@@href[1]{\endgroup#1\@@endlink}%
\providecommand \@sanitize@url [0]{\catcode `\\12\catcode `\$12\catcode
  `\&12\catcode `\#12\catcode `\^12\catcode `\_12\catcode `\%12\relax}%
\providecommand \@@startlink[1]{}%
\providecommand \@@endlink[0]{}%
\providecommand \url  [0]{\begingroup\@sanitize@url \@url }%
\providecommand \@url [1]{\endgroup\@href {#1}{\urlprefix }}%
\providecommand \urlprefix  [0]{URL }%
\providecommand \Eprint [0]{\href }%
\providecommand \doibase [0]{http://dx.doi.org/}%
\providecommand \selectlanguage [0]{\@gobble}%
\providecommand \bibinfo  [0]{\@secondoftwo}%
\providecommand \bibfield  [0]{\@secondoftwo}%
\providecommand \translation [1]{[#1]}%
\providecommand \BibitemOpen [0]{}%
\providecommand \bibitemStop [0]{}%
\providecommand \bibitemNoStop [0]{.\EOS\space}%
\providecommand \EOS [0]{\spacefactor3000\relax}%
\providecommand \BibitemShut  [1]{\csname bibitem#1\endcsname}%
\let\auto@bib@innerbib\@empty
\bibitem [{\citenamefont {Gor'kov}\ and\ \citenamefont
  {Rashba}(2001)}]{Gorkov_PRL_2001}%
  \BibitemOpen
  \bibfield  {author} {\bibinfo {author} {\bibfnamefont {Lev~P.}\ \bibnamefont
  {Gor'kov}}\ and\ \bibinfo {author} {\bibfnamefont {Emmanuel~I.}\ \bibnamefont
  {Rashba}},\ }\bibfield  {title} {\enquote {\bibinfo {title} {{Superconducting
  2D System with Lifted Spin Degeneracy: Mixed Singlet-Triplet State}},}\
  }\href {\doibase 10.1103/PhysRevLett.87.037004} {\bibfield  {journal}
  {\bibinfo  {journal} {Phys. Rev. Lett.}\ }\textbf {\bibinfo {volume} {87}},\
  \bibinfo {pages} {037004} (\bibinfo {year} {2001})}\BibitemShut {NoStop}%
\bibitem [{\citenamefont {Sato}\ and\ \citenamefont
  {Fujimoto}(2009)}]{Sato_PRB_2009}%
  \BibitemOpen
  \bibfield  {author} {\bibinfo {author} {\bibfnamefont {Masatoshi}\
  \bibnamefont {Sato}}\ and\ \bibinfo {author} {\bibfnamefont {Satoshi}\
  \bibnamefont {Fujimoto}},\ }\bibfield  {title} {\enquote {\bibinfo {title}
  {{Topological phases of noncentrosymmetric superconductors: Edge states,
  Majorana fermions, and non-Abelian statistics}},}\ }\href {\doibase
  10.1103/PhysRevB.79.094504} {\bibfield  {journal} {\bibinfo  {journal} {Phys.
  Rev. B}\ }\textbf {\bibinfo {volume} {79}},\ \bibinfo {pages} {094504}
  (\bibinfo {year} {2009})}\BibitemShut {NoStop}%
\bibitem [{\citenamefont {Mizushima}\ \emph {et~al.}(2014)\citenamefont
  {Mizushima}, \citenamefont {Yamakage}, \citenamefont {Sato},\ and\
  \citenamefont {Tanaka}}]{Mizushima_PRB_2014}%
  \BibitemOpen
  \bibfield  {author} {\bibinfo {author} {\bibfnamefont {Takeshi}\ \bibnamefont
  {Mizushima}}, \bibinfo {author} {\bibfnamefont {Ai}~\bibnamefont {Yamakage}},
  \bibinfo {author} {\bibfnamefont {Masatoshi}\ \bibnamefont {Sato}}, \ and\
  \bibinfo {author} {\bibfnamefont {Yukio}\ \bibnamefont {Tanaka}},\ }\bibfield
   {title} {\enquote {\bibinfo {title} {Dirac-fermion-induced parity mixing in
  superconducting topological insulators},}\ }\href {\doibase
  10.1103/PhysRevB.90.184516} {\bibfield  {journal} {\bibinfo  {journal} {Phys.
  Rev. B}\ }\textbf {\bibinfo {volume} {90}},\ \bibinfo {pages} {184516}
  (\bibinfo {year} {2014})}\BibitemShut {NoStop}%
\bibitem [{\citenamefont {Yao}\ and\ \citenamefont
  {Yang}(2015)}]{Yao_PRB_2015}%
  \BibitemOpen
  \bibfield  {author} {\bibinfo {author} {\bibfnamefont {Hong}\ \bibnamefont
  {Yao}}\ and\ \bibinfo {author} {\bibfnamefont {Fan}\ \bibnamefont {Yang}},\
  }\bibfield  {title} {\enquote {\bibinfo {title} {{Topological odd-parity
  superconductivity at type-II two-dimensional van Hove singularities}},}\
  }\href {\doibase 10.1103/PhysRevB.92.035132} {\bibfield  {journal} {\bibinfo
  {journal} {Phys. Rev. B}\ }\textbf {\bibinfo {volume} {92}},\ \bibinfo
  {pages} {035132} (\bibinfo {year} {2015})}\BibitemShut {NoStop}%
\bibitem [{\citenamefont {Lu}\ and\ \citenamefont
  {S\'en\'echal}(2018)}]{Lu_PRB_2018}%
  \BibitemOpen
  \bibfield  {author} {\bibinfo {author} {\bibfnamefont {Xiancong}\
  \bibnamefont {Lu}}\ and\ \bibinfo {author} {\bibfnamefont {D.}~\bibnamefont
  {S\'en\'echal}},\ }\bibfield  {title} {\enquote {\bibinfo {title}
  {{Parity-mixing superconducting phase in the Rashba-Hubbard model and its
  topological properties from dynamical mean-field theory}},}\ }\href {\doibase
  10.1103/PhysRevB.98.245118} {\bibfield  {journal} {\bibinfo  {journal} {Phys.
  Rev. B}\ }\textbf {\bibinfo {volume} {98}},\ \bibinfo {pages} {245118}
  (\bibinfo {year} {2018})}\BibitemShut {NoStop}%
\bibitem [{\citenamefont {Trott}\ and\ \citenamefont
  {Hooley}(2020)}]{Trott_PRR_2020}%
  \BibitemOpen
  \bibfield  {author} {\bibinfo {author} {\bibfnamefont {Matthew~J.}\
  \bibnamefont {Trott}}\ and\ \bibinfo {author} {\bibfnamefont {Chris~A.}\
  \bibnamefont {Hooley}},\ }\bibfield  {title} {\enquote {\bibinfo {title}
  {{Mixed-parity superconductivity near Lifshitz transitions in strongly
  spin-orbit-coupled metals}},}\ }\href {\doibase
  10.1103/PhysRevResearch.2.013106} {\bibfield  {journal} {\bibinfo  {journal}
  {Phys. Rev. Research}\ }\textbf {\bibinfo {volume} {2}},\ \bibinfo {pages}
  {013106} (\bibinfo {year} {2020})}\BibitemShut {NoStop}%
\bibitem [{\citenamefont {Mourik}\ \emph {et~al.}(2012)\citenamefont {Mourik},
  \citenamefont {Zuo}, \citenamefont {Frolov}, \citenamefont {Plissard},
  \citenamefont {Bakkers},\ and\ \citenamefont
  {Kouwenhoven}}]{Mourik_Science_2012}%
  \BibitemOpen
  \bibfield  {author} {\bibinfo {author} {\bibfnamefont {V.}~\bibnamefont
  {Mourik}}, \bibinfo {author} {\bibfnamefont {K.}~\bibnamefont {Zuo}},
  \bibinfo {author} {\bibfnamefont {S.~M.}\ \bibnamefont {Frolov}}, \bibinfo
  {author} {\bibfnamefont {S.~R.}\ \bibnamefont {Plissard}}, \bibinfo {author}
  {\bibfnamefont {E.~P. A.~M.}\ \bibnamefont {Bakkers}}, \ and\ \bibinfo
  {author} {\bibfnamefont {L.~P.}\ \bibnamefont {Kouwenhoven}},\ }\bibfield
  {title} {\enquote {\bibinfo {title} {{Signatures of Majorana Fermions in
  Hybrid Superconductor-Semiconductor Nanowire Devices}},}\ }\href {\doibase
  10.1126/science.1222360} {\bibfield  {journal} {\bibinfo  {journal}
  {Science}\ }\textbf {\bibinfo {volume} {336}},\ \bibinfo {pages} {1003--1007}
  (\bibinfo {year} {2012})}\BibitemShut {NoStop}%
\bibitem [{\citenamefont {Zhang}\ \emph
  {et~al.}(2018{\natexlab{a}})\citenamefont {Zhang}, \citenamefont {Liu},
  \citenamefont {Gazibegovic}, \citenamefont {Xu}, \citenamefont {Logan},
  \citenamefont {Wang}, \citenamefont {Loo}, \citenamefont {Bommer},
  \citenamefont {Moor}, \citenamefont {Car}, \citenamefont {Veld},
  \citenamefont {Veldhoven}, \citenamefont {Koelling}, \citenamefont
  {Verheijen}, \citenamefont {Pendharkar}, \citenamefont {Pennachio},
  \citenamefont {Shojaei}, \citenamefont {Lee}, \citenamefont {Palmstrøm},
  \citenamefont {Bakkers}, \citenamefont {Sarma},\ and\ \citenamefont
  {Kouwenhoven}}]{Zhang_Nature_2018}%
  \BibitemOpen
  \bibfield  {author} {\bibinfo {author} {\bibfnamefont {Hao}\ \bibnamefont
  {Zhang}}, \bibinfo {author} {\bibfnamefont {Chun-Xiao}\ \bibnamefont {Liu}},
  \bibinfo {author} {\bibfnamefont {Sasa}\ \bibnamefont {Gazibegovic}},
  \bibinfo {author} {\bibfnamefont {Di}~\bibnamefont {Xu}}, \bibinfo {author}
  {\bibfnamefont {John~A}\ \bibnamefont {Logan}}, \bibinfo {author}
  {\bibfnamefont {Guanzhong}\ \bibnamefont {Wang}}, \bibinfo {author}
  {\bibfnamefont {Nick~van}\ \bibnamefont {Loo}}, \bibinfo {author}
  {\bibfnamefont {Jouri~DS}\ \bibnamefont {Bommer}}, \bibinfo {author}
  {\bibfnamefont {Michiel WA~de}\ \bibnamefont {Moor}}, \bibinfo {author}
  {\bibfnamefont {Diana}\ \bibnamefont {Car}}, \bibinfo {author} {\bibfnamefont
  {Roy LM~het}\ \bibnamefont {Veld}}, \bibinfo {author} {\bibfnamefont {Petrus
  J~van}\ \bibnamefont {Veldhoven}}, \bibinfo {author} {\bibfnamefont
  {Sebastian}\ \bibnamefont {Koelling}}, \bibinfo {author} {\bibfnamefont
  {Marcel~A}\ \bibnamefont {Verheijen}}, \bibinfo {author} {\bibfnamefont
  {Mihir}\ \bibnamefont {Pendharkar}}, \bibinfo {author} {\bibfnamefont
  {Daniel~J}\ \bibnamefont {Pennachio}}, \bibinfo {author} {\bibfnamefont
  {Borzoyeh}\ \bibnamefont {Shojaei}}, \bibinfo {author} {\bibfnamefont {Joon}\
  \bibnamefont {Lee}}, \bibinfo {author} {\bibfnamefont {Chris~J}\ \bibnamefont
  {Palmstrøm}}, \bibinfo {author} {\bibfnamefont {Erik}\ \bibnamefont
  {Bakkers}}, \bibinfo {author} {\bibfnamefont {Das~S}\ \bibnamefont {Sarma}},
  \ and\ \bibinfo {author} {\bibfnamefont {Leo~P}\ \bibnamefont
  {Kouwenhoven}},\ }\bibfield  {title} {\enquote {\bibinfo {title} {{Quantized
  Majorana conductance}},}\ }\href {\doibase 10.1038/nature26142} {\bibfield
  {journal} {\bibinfo  {journal} {Nature}\ }\textbf {\bibinfo {volume} {556}},\
  \bibinfo {pages} {74} (\bibinfo {year} {2018}{\natexlab{a}})}\BibitemShut
  {NoStop}%
\bibitem [{\citenamefont {Lutchyn}\ \emph {et~al.}(2018)\citenamefont
  {Lutchyn}, \citenamefont {Bakkers}, \citenamefont {Kouwenhoven},
  \citenamefont {Krogstrup}, \citenamefont {Marcus},\ and\ \citenamefont
  {Oreg}}]{Lutchyn_NRM_2018}%
  \BibitemOpen
  \bibfield  {author} {\bibinfo {author} {\bibfnamefont {RM}~\bibnamefont
  {Lutchyn}}, \bibinfo {author} {\bibfnamefont {EPAM}\ \bibnamefont {Bakkers}},
  \bibinfo {author} {\bibfnamefont {LP}~\bibnamefont {Kouwenhoven}}, \bibinfo
  {author} {\bibfnamefont {P}~\bibnamefont {Krogstrup}}, \bibinfo {author}
  {\bibfnamefont {CM}~\bibnamefont {Marcus}}, \ and\ \bibinfo {author}
  {\bibfnamefont {Y}~\bibnamefont {Oreg}},\ }\bibfield  {title} {\enquote
  {\bibinfo {title} {{Majorana zero modes in superconductor–semiconductor
  heterostructures}},}\ }\href {\doibase 10.1038/s41578-018-0003-1} {\bibfield
  {journal} {\bibinfo  {journal} {Nature Reviews Materials}\ }\textbf {\bibinfo
  {volume} {3}},\ \bibinfo {pages} {52--68} (\bibinfo {year}
  {2018})}\BibitemShut {NoStop}%
\bibitem [{\citenamefont {Manna}\ \emph {et~al.}(2020)\citenamefont {Manna},
  \citenamefont {Wei}, \citenamefont {Xie}, \citenamefont {Law}, \citenamefont
  {Lee},\ and\ \citenamefont {Moodera}}]{Manna_PNAS_2020}%
  \BibitemOpen
  \bibfield  {author} {\bibinfo {author} {\bibfnamefont {Sujit}\ \bibnamefont
  {Manna}}, \bibinfo {author} {\bibfnamefont {Peng}\ \bibnamefont {Wei}},
  \bibinfo {author} {\bibfnamefont {Yingming}\ \bibnamefont {Xie}}, \bibinfo
  {author} {\bibfnamefont {Kam~Tuen}\ \bibnamefont {Law}}, \bibinfo {author}
  {\bibfnamefont {Patrick~A.}\ \bibnamefont {Lee}}, \ and\ \bibinfo {author}
  {\bibfnamefont {Jagadeesh~S.}\ \bibnamefont {Moodera}},\ }\bibfield  {title}
  {\enquote {\bibinfo {title} {{Signature of a pair of Majorana zero modes in
  superconducting gold surface states}},}\ }\href {\doibase
  10.1073/pnas.1919753117} {\bibfield  {journal} {\bibinfo  {journal} {Proc.
  Natl. Acad. Sci.}\ }\textbf {\bibinfo {volume} {117}},\ \bibinfo {pages}
  {8775--8782} (\bibinfo {year} {2020})}\BibitemShut {NoStop}%
\bibitem [{\citenamefont {Kezilebieke}\ \emph {et~al.}(2020)\citenamefont
  {Kezilebieke}, \citenamefont {Huda}, \citenamefont {Vaňo}, \citenamefont
  {Aapro}, \citenamefont {Ganguli}, \citenamefont {Silveira}, \citenamefont
  {Głodzik}, \citenamefont {Foster}, \citenamefont {Ojanen},\ and\
  \citenamefont {Liljeroth}}]{Kezilebieke_Nature_2020}%
  \BibitemOpen
  \bibfield  {author} {\bibinfo {author} {\bibfnamefont {Shawulienu}\
  \bibnamefont {Kezilebieke}}, \bibinfo {author} {\bibfnamefont {Md~Nurul}\
  \bibnamefont {Huda}}, \bibinfo {author} {\bibfnamefont {Viliam}\ \bibnamefont
  {Vaňo}}, \bibinfo {author} {\bibfnamefont {Markus}\ \bibnamefont {Aapro}},
  \bibinfo {author} {\bibfnamefont {Somesh~C.}\ \bibnamefont {Ganguli}},
  \bibinfo {author} {\bibfnamefont {Orlando~J.}\ \bibnamefont {Silveira}},
  \bibinfo {author} {\bibfnamefont {Szczepan}\ \bibnamefont {Głodzik}},
  \bibinfo {author} {\bibfnamefont {Adam~S.}\ \bibnamefont {Foster}}, \bibinfo
  {author} {\bibfnamefont {Teemu}\ \bibnamefont {Ojanen}}, \ and\ \bibinfo
  {author} {\bibfnamefont {Peter}\ \bibnamefont {Liljeroth}},\ }\bibfield
  {title} {\enquote {\bibinfo {title} {{Topological superconductivity in a van
  der Waals heterostructure}},}\ }\href {\doibase 10.1038/s41586-020-2989-y}
  {\bibfield  {journal} {\bibinfo  {journal} {Nature}\ }\textbf {\bibinfo
  {volume} {588}},\ \bibinfo {pages} {424--428} (\bibinfo {year}
  {2020})}\BibitemShut {NoStop}%
\bibitem [{\citenamefont {Zhang}\ \emph
  {et~al.}(2018{\natexlab{b}})\citenamefont {Zhang}, \citenamefont {Yaji},
  \citenamefont {Hashimoto}, \citenamefont {Ota}, \citenamefont {Kondo},
  \citenamefont {Okazaki}, \citenamefont {Wang}, \citenamefont {Wen},
  \citenamefont {Gu}, \citenamefont {Ding},\ and\ \citenamefont
  {Shin}}]{Zhang_Science_2018}%
  \BibitemOpen
  \bibfield  {author} {\bibinfo {author} {\bibfnamefont {Peng}\ \bibnamefont
  {Zhang}}, \bibinfo {author} {\bibfnamefont {Koichiro}\ \bibnamefont {Yaji}},
  \bibinfo {author} {\bibfnamefont {Takahiro}\ \bibnamefont {Hashimoto}},
  \bibinfo {author} {\bibfnamefont {Yuichi}\ \bibnamefont {Ota}}, \bibinfo
  {author} {\bibfnamefont {Takeshi}\ \bibnamefont {Kondo}}, \bibinfo {author}
  {\bibfnamefont {Kozo}\ \bibnamefont {Okazaki}}, \bibinfo {author}
  {\bibfnamefont {Zhijun}\ \bibnamefont {Wang}}, \bibinfo {author}
  {\bibfnamefont {Jinsheng}\ \bibnamefont {Wen}}, \bibinfo {author}
  {\bibfnamefont {G.~D.}\ \bibnamefont {Gu}}, \bibinfo {author} {\bibfnamefont
  {Hong}\ \bibnamefont {Ding}}, \ and\ \bibinfo {author} {\bibfnamefont {Shik}\
  \bibnamefont {Shin}},\ }\bibfield  {title} {\enquote {\bibinfo {title}
  {Observation of topological superconductivity on the surface of an iron-based
  superconductor},}\ }\href
  {https://www.science.org/doi/abs/10.1126/science.aan4596} {\bibfield
  {journal} {\bibinfo  {journal} {Science}\ }\textbf {\bibinfo {volume}
  {360}},\ \bibinfo {pages} {182--186} (\bibinfo {year}
  {2018}{\natexlab{b}})}\BibitemShut {NoStop}%
\bibitem [{\citenamefont {Wang}\ \emph {et~al.}(2018)\citenamefont {Wang},
  \citenamefont {Kong}, \citenamefont {Fan}, \citenamefont {Chen},
  \citenamefont {Zhu}, \citenamefont {Liu}, \citenamefont {Cao}, \citenamefont
  {Sun}, \citenamefont {Du}, \citenamefont {Schneeloch}, \citenamefont {Zhong},
  \citenamefont {Gu}, \citenamefont {Fu}, \citenamefont {Ding},\ and\
  \citenamefont {Gao}}]{Wang_Science_2018}%
  \BibitemOpen
  \bibfield  {author} {\bibinfo {author} {\bibfnamefont {Dongfei}\ \bibnamefont
  {Wang}}, \bibinfo {author} {\bibfnamefont {Lingyuan}\ \bibnamefont {Kong}},
  \bibinfo {author} {\bibfnamefont {Peng}\ \bibnamefont {Fan}}, \bibinfo
  {author} {\bibfnamefont {Hui}\ \bibnamefont {Chen}}, \bibinfo {author}
  {\bibfnamefont {Shiyu}\ \bibnamefont {Zhu}}, \bibinfo {author} {\bibfnamefont
  {Wenyao}\ \bibnamefont {Liu}}, \bibinfo {author} {\bibfnamefont
  {Lu}~\bibnamefont {Cao}}, \bibinfo {author} {\bibfnamefont {Yujie}\
  \bibnamefont {Sun}}, \bibinfo {author} {\bibfnamefont {Shixuan}\ \bibnamefont
  {Du}}, \bibinfo {author} {\bibfnamefont {John}\ \bibnamefont {Schneeloch}},
  \bibinfo {author} {\bibfnamefont {Ruidan}\ \bibnamefont {Zhong}}, \bibinfo
  {author} {\bibfnamefont {Genda}\ \bibnamefont {Gu}}, \bibinfo {author}
  {\bibfnamefont {Liang}\ \bibnamefont {Fu}}, \bibinfo {author} {\bibfnamefont
  {Hong}\ \bibnamefont {Ding}}, \ and\ \bibinfo {author} {\bibfnamefont
  {Hong-Jun}\ \bibnamefont {Gao}},\ }\bibfield  {title} {\enquote {\bibinfo
  {title} {{Evidence for Majorana bound states in an iron-based
  superconductor}},}\ }\href
  {https://science.sciencemag.org/content/362/6412/333} {\bibfield  {journal}
  {\bibinfo  {journal} {Science}\ }\textbf {\bibinfo {volume} {362}},\ \bibinfo
  {pages} {333--335} (\bibinfo {year} {2018})}\BibitemShut {NoStop}%
\bibitem [{\citenamefont {Machida}\ \emph {et~al.}(2019)\citenamefont
  {Machida}, \citenamefont {Sun}, \citenamefont {Pyon}, \citenamefont {Takeda},
  \citenamefont {Kohsaka}, \citenamefont {Hanaguri}, \citenamefont {Sasagawa},\
  and\ \citenamefont {Tamegai}}]{Machida_NM_2019}%
  \BibitemOpen
  \bibfield  {author} {\bibinfo {author} {\bibfnamefont {T.}~\bibnamefont
  {Machida}}, \bibinfo {author} {\bibfnamefont {Y.}~\bibnamefont {Sun}},
  \bibinfo {author} {\bibfnamefont {S.}~\bibnamefont {Pyon}}, \bibinfo {author}
  {\bibfnamefont {S.}~\bibnamefont {Takeda}}, \bibinfo {author} {\bibfnamefont
  {Y.}~\bibnamefont {Kohsaka}}, \bibinfo {author} {\bibfnamefont
  {T.}~\bibnamefont {Hanaguri}}, \bibinfo {author} {\bibfnamefont
  {T.}~\bibnamefont {Sasagawa}}, \ and\ \bibinfo {author} {\bibfnamefont
  {T.}~\bibnamefont {Tamegai}},\ }\bibfield  {title} {\enquote {\bibinfo
  {title} {{Zero-energy vortex bound state in the superconducting topological
  surface state of Fe(Se,Te)}},}\ }\href {\doibase 10.1038/s41563-019-0397-1}
  {\bibfield  {journal} {\bibinfo  {journal} {Nature Materials}\ }\textbf
  {\bibinfo {volume} {18}},\ \bibinfo {pages} {811--815} (\bibinfo {year}
  {2019})}\BibitemShut {NoStop}%
\bibitem [{\citenamefont {Iwaya}\ \emph {et~al.}(2017)\citenamefont {Iwaya},
  \citenamefont {Kohsaka}, \citenamefont {Okawa}, \citenamefont {Machida},
  \citenamefont {Bahramy}, \citenamefont {Hanaguri},\ and\ \citenamefont
  {Sasagawa}}]{Iwaya_NC_2017}%
  \BibitemOpen
  \bibfield  {author} {\bibinfo {author} {\bibfnamefont {K.}~\bibnamefont
  {Iwaya}}, \bibinfo {author} {\bibfnamefont {Y.}~\bibnamefont {Kohsaka}},
  \bibinfo {author} {\bibfnamefont {K.}~\bibnamefont {Okawa}}, \bibinfo
  {author} {\bibfnamefont {T.}~\bibnamefont {Machida}}, \bibinfo {author}
  {\bibfnamefont {M.~S.}\ \bibnamefont {Bahramy}}, \bibinfo {author}
  {\bibfnamefont {T.}~\bibnamefont {Hanaguri}}, \ and\ \bibinfo {author}
  {\bibfnamefont {T.}~\bibnamefont {Sasagawa}},\ }\bibfield  {title} {\enquote
  {\bibinfo {title} {{Full-gap superconductivity in spin-polarised surface
  states of topological semimetal $\beta$-PdBi$_2$}},}\ }\href {\doibase
  10.1038/s41467-017-01209-9} {\bibfield  {journal} {\bibinfo  {journal}
  {Nature Communications}\ }\textbf {\bibinfo {volume} {8}},\ \bibinfo {pages}
  {976} (\bibinfo {year} {2017})}\BibitemShut {NoStop}%
\bibitem [{\citenamefont {Guan}\ \emph {et~al.}(2016)\citenamefont {Guan},
  \citenamefont {Chen}, \citenamefont {Chu}, \citenamefont {Sankar},
  \citenamefont {Chou}, \citenamefont {Jeng}, \citenamefont {Chang},\ and\
  \citenamefont {Chuang}}]{Guan_SAvd_2018}%
  \BibitemOpen
  \bibfield  {author} {\bibinfo {author} {\bibfnamefont {Syu-You}\ \bibnamefont
  {Guan}}, \bibinfo {author} {\bibfnamefont {Peng-Jen}\ \bibnamefont {Chen}},
  \bibinfo {author} {\bibfnamefont {Ming-Wen}\ \bibnamefont {Chu}}, \bibinfo
  {author} {\bibfnamefont {Raman}\ \bibnamefont {Sankar}}, \bibinfo {author}
  {\bibfnamefont {Fangcheng}\ \bibnamefont {Chou}}, \bibinfo {author}
  {\bibfnamefont {Horng-Tay}\ \bibnamefont {Jeng}}, \bibinfo {author}
  {\bibfnamefont {Chia-Seng}\ \bibnamefont {Chang}}, \ and\ \bibinfo {author}
  {\bibfnamefont {Tien-Ming}\ \bibnamefont {Chuang}},\ }\bibfield  {title}
  {\enquote {\bibinfo {title} {{Superconducting topological surface states in
  the noncentrosymmetric bulk superconductor PbTaSe$_2$}},}\ }\href {\doibase
  10.1126/sciadv.1600894} {\bibfield  {journal} {\bibinfo  {journal} {Science
  Advances}\ }\textbf {\bibinfo {volume} {2}},\ \bibinfo {pages} {e1600894}
  (\bibinfo {year} {2016})}\BibitemShut {NoStop}%
\bibitem [{\citenamefont {Zhang}\ \emph {et~al.}(2010)\citenamefont {Zhang},
  \citenamefont {Cheng}, \citenamefont {Li}, \citenamefont {Sun}, \citenamefont
  {Wang}, \citenamefont {Zhu}, \citenamefont {He}, \citenamefont {Wang},
  \citenamefont {Ma}, \citenamefont {Chen}, \citenamefont {Wang}, \citenamefont
  {Liu}, \citenamefont {Lin}, \citenamefont {Jia},\ and\ \citenamefont
  {Xue}}]{Zhang_NP_2010}%
  \BibitemOpen
  \bibfield  {author} {\bibinfo {author} {\bibfnamefont {Tong}\ \bibnamefont
  {Zhang}}, \bibinfo {author} {\bibfnamefont {Peng}\ \bibnamefont {Cheng}},
  \bibinfo {author} {\bibfnamefont {Wen-Juan}\ \bibnamefont {Li}}, \bibinfo
  {author} {\bibfnamefont {Yu-Jie}\ \bibnamefont {Sun}}, \bibinfo {author}
  {\bibfnamefont {Guang}\ \bibnamefont {Wang}}, \bibinfo {author}
  {\bibfnamefont {Xie-Gang}\ \bibnamefont {Zhu}}, \bibinfo {author}
  {\bibfnamefont {Ke}~\bibnamefont {He}}, \bibinfo {author} {\bibfnamefont
  {Lili}\ \bibnamefont {Wang}}, \bibinfo {author} {\bibfnamefont {Xucun}\
  \bibnamefont {Ma}}, \bibinfo {author} {\bibfnamefont {Xi}~\bibnamefont
  {Chen}}, \bibinfo {author} {\bibfnamefont {Yayu}\ \bibnamefont {Wang}},
  \bibinfo {author} {\bibfnamefont {Ying}\ \bibnamefont {Liu}}, \bibinfo
  {author} {\bibfnamefont {Hai-Qing}\ \bibnamefont {Lin}}, \bibinfo {author}
  {\bibfnamefont {Jin-Feng}\ \bibnamefont {Jia}}, \ and\ \bibinfo {author}
  {\bibfnamefont {Qi-Kun}\ \bibnamefont {Xue}},\ }\bibfield  {title} {\enquote
  {\bibinfo {title} {{Superconductivity in one-atomic-layer metal films grown
  on Si(111)}},}\ }\href {\doibase 10.1038/nphys1499} {\bibfield  {journal}
  {\bibinfo  {journal} {Nature Physics}\ }\textbf {\bibinfo {volume} {6}},\
  \bibinfo {pages} {104--108} (\bibinfo {year} {2010})}\BibitemShut {NoStop}%
\bibitem [{\citenamefont {Yoshizawa}\ \emph {et~al.}(2014)\citenamefont
  {Yoshizawa}, \citenamefont {Kim}, \citenamefont {Kawakami}, \citenamefont
  {Nagai}, \citenamefont {Nakayama}, \citenamefont {Hu}, \citenamefont
  {Hasegawa},\ and\ \citenamefont {Uchihashi}}]{Yoshizawa_PRL_2014}%
  \BibitemOpen
  \bibfield  {author} {\bibinfo {author} {\bibfnamefont {Shunsuke}\
  \bibnamefont {Yoshizawa}}, \bibinfo {author} {\bibfnamefont {Howon}\
  \bibnamefont {Kim}}, \bibinfo {author} {\bibfnamefont {Takuto}\ \bibnamefont
  {Kawakami}}, \bibinfo {author} {\bibfnamefont {Yuki}\ \bibnamefont {Nagai}},
  \bibinfo {author} {\bibfnamefont {Tomonobu}\ \bibnamefont {Nakayama}},
  \bibinfo {author} {\bibfnamefont {Xiao}\ \bibnamefont {Hu}}, \bibinfo
  {author} {\bibfnamefont {Yukio}\ \bibnamefont {Hasegawa}}, \ and\ \bibinfo
  {author} {\bibfnamefont {Takashi}\ \bibnamefont {Uchihashi}},\ }\bibfield
  {title} {\enquote {\bibinfo {title} {{Imaging Josephson Vortices on the
  Surface Superconductor
  $\mathrm{Si}(111)\text{\ensuremath{-}}(\sqrt{7}\ifmmode\times\else\texttimes\fi{}\sqrt{3})\text{\ensuremath{-}}\mathrm{In}$
  using a Scanning Tunneling Microscope}},}\ }\href {\doibase
  10.1103/PhysRevLett.113.247004} {\bibfield  {journal} {\bibinfo  {journal}
  {Phys. Rev. Lett.}\ }\textbf {\bibinfo {volume} {113}},\ \bibinfo {pages}
  {247004} (\bibinfo {year} {2014})}\BibitemShut {NoStop}%
\bibitem [{\citenamefont {Yoshizawa}\ \emph {et~al.}(2021)\citenamefont
  {Yoshizawa}, \citenamefont {Kobayashi}, \citenamefont {Nakata}, \citenamefont
  {Yaji}, \citenamefont {Yokota}, \citenamefont {Komori}, \citenamefont {Shin},
  \citenamefont {Sakamoto},\ and\ \citenamefont
  {Uchihashi}}]{Yoshizawa_NC_2021}%
  \BibitemOpen
  \bibfield  {author} {\bibinfo {author} {\bibfnamefont {Shunsuke}\
  \bibnamefont {Yoshizawa}}, \bibinfo {author} {\bibfnamefont {Takahiro}\
  \bibnamefont {Kobayashi}}, \bibinfo {author} {\bibfnamefont {Yoshitaka}\
  \bibnamefont {Nakata}}, \bibinfo {author} {\bibfnamefont {Koichiro}\
  \bibnamefont {Yaji}}, \bibinfo {author} {\bibfnamefont {Kenta}\ \bibnamefont
  {Yokota}}, \bibinfo {author} {\bibfnamefont {Fumio}\ \bibnamefont {Komori}},
  \bibinfo {author} {\bibfnamefont {Shik}\ \bibnamefont {Shin}}, \bibinfo
  {author} {\bibfnamefont {Kazuyuki}\ \bibnamefont {Sakamoto}}, \ and\ \bibinfo
  {author} {\bibfnamefont {Takashi}\ \bibnamefont {Uchihashi}},\ }\bibfield
  {title} {\enquote {\bibinfo {title} {{Atomic-layer Rashba-type superconductor
  protected by dynamic spin-momentum locking}},}\ }\href {\doibase
  10.1038/s41467-021-21642-1} {\bibfield  {journal} {\bibinfo  {journal}
  {Nature Communications}\ }\textbf {\bibinfo {volume} {12}},\ \bibinfo {pages}
  {1462} (\bibinfo {year} {2021})},\ \Eprint {http://arxiv.org/abs/2103.07143}
  {2103.07143} \BibitemShut {NoStop}%
\bibitem [{\citenamefont {Matetskiy}\ \emph {et~al.}(2015)\citenamefont
  {Matetskiy}, \citenamefont {Ichinokura}, \citenamefont {Bondarenko},
  \citenamefont {Tupchaya}, \citenamefont {Gruznev}, \citenamefont {Zotov},
  \citenamefont {Saranin}, \citenamefont {Hobara}, \citenamefont {Takayama},\
  and\ \citenamefont {Hasegawa}}]{Matetskiy_PRL_2015}%
  \BibitemOpen
  \bibfield  {author} {\bibinfo {author} {\bibfnamefont {A.~V.}\ \bibnamefont
  {Matetskiy}}, \bibinfo {author} {\bibfnamefont {S.}~\bibnamefont
  {Ichinokura}}, \bibinfo {author} {\bibfnamefont {L.~V.}\ \bibnamefont
  {Bondarenko}}, \bibinfo {author} {\bibfnamefont {A.~Y.}\ \bibnamefont
  {Tupchaya}}, \bibinfo {author} {\bibfnamefont {D.~V.}\ \bibnamefont
  {Gruznev}}, \bibinfo {author} {\bibfnamefont {A.~V.}\ \bibnamefont {Zotov}},
  \bibinfo {author} {\bibfnamefont {A.~A.}\ \bibnamefont {Saranin}}, \bibinfo
  {author} {\bibfnamefont {R.}~\bibnamefont {Hobara}}, \bibinfo {author}
  {\bibfnamefont {A.}~\bibnamefont {Takayama}}, \ and\ \bibinfo {author}
  {\bibfnamefont {S.}~\bibnamefont {Hasegawa}},\ }\bibfield  {title} {\enquote
  {\bibinfo {title} {{Two-Dimensional Superconductor with a Giant Rashba
  Effect: One-Atom-Layer Tl-Pb Compound on Si(111)}},}\ }\href {\doibase
  10.1103/PhysRevLett.115.147003} {\bibfield  {journal} {\bibinfo  {journal}
  {Phys. Rev. Lett.}\ }\textbf {\bibinfo {volume} {115}},\ \bibinfo {pages}
  {147003} (\bibinfo {year} {2015})}\BibitemShut {NoStop}%
\bibitem [{\citenamefont {Gruznev}\ \emph {et~al.}(2014)\citenamefont
  {Gruznev}, \citenamefont {Bondarenko}, \citenamefont {Matetskiy},
  \citenamefont {Yakovlev}, \citenamefont {Tupchaya}, \citenamefont {Eremeev},
  \citenamefont {Chulkov}, \citenamefont {Chou}, \citenamefont {Wei},
  \citenamefont {Lai}, \citenamefont {Wang}, \citenamefont {Zotov},\ and\
  \citenamefont {Saranin}}]{Gruznev_SRep_2014}%
  \BibitemOpen
  \bibfield  {author} {\bibinfo {author} {\bibfnamefont {Dimitry~V.}\
  \bibnamefont {Gruznev}}, \bibinfo {author} {\bibfnamefont {Leonid~V.}\
  \bibnamefont {Bondarenko}}, \bibinfo {author} {\bibfnamefont {Andrey~V.}\
  \bibnamefont {Matetskiy}}, \bibinfo {author} {\bibfnamefont {Alexey~A.}\
  \bibnamefont {Yakovlev}}, \bibinfo {author} {\bibfnamefont {Alexandra~Y.}\
  \bibnamefont {Tupchaya}}, \bibinfo {author} {\bibfnamefont {Sergey~V.}\
  \bibnamefont {Eremeev}}, \bibinfo {author} {\bibfnamefont {Evgeniy~V.}\
  \bibnamefont {Chulkov}}, \bibinfo {author} {\bibfnamefont {Jyh-Pin}\
  \bibnamefont {Chou}}, \bibinfo {author} {\bibfnamefont {Ching-Ming}\
  \bibnamefont {Wei}}, \bibinfo {author} {\bibfnamefont {Ming-Yu}\ \bibnamefont
  {Lai}}, \bibinfo {author} {\bibfnamefont {Yuh-Lin}\ \bibnamefont {Wang}},
  \bibinfo {author} {\bibfnamefont {Andrey~V.}\ \bibnamefont {Zotov}}, \ and\
  \bibinfo {author} {\bibfnamefont {Alexander~A.}\ \bibnamefont {Saranin}},\
  }\bibfield  {title} {\enquote {\bibinfo {title} {{A Strategy to Create
  Spin-Split Metallic Bands on Silicon Using a Dense Alloy Layer}},}\ }\href
  {\doibase 10.1038/srep04742} {\bibfield  {journal} {\bibinfo  {journal}
  {Scientific Reports}\ }\textbf {\bibinfo {volume} {4}},\ \bibinfo {pages}
  {4742} (\bibinfo {year} {2014})}\BibitemShut {NoStop}%
\bibitem [{\citenamefont {Nakamura}\ \emph {et~al.}(2018)\citenamefont
  {Nakamura}, \citenamefont {Kim}, \citenamefont {Ichinokura}, \citenamefont
  {Takayama}, \citenamefont {Zotov}, \citenamefont {Saranin}, \citenamefont
  {Hasegawa},\ and\ \citenamefont {Hasegawa}}]{Nakamura_PRB_2018}%
  \BibitemOpen
  \bibfield  {author} {\bibinfo {author} {\bibfnamefont {T.}~\bibnamefont
  {Nakamura}}, \bibinfo {author} {\bibfnamefont {H.}~\bibnamefont {Kim}},
  \bibinfo {author} {\bibfnamefont {S.}~\bibnamefont {Ichinokura}}, \bibinfo
  {author} {\bibfnamefont {A.}~\bibnamefont {Takayama}}, \bibinfo {author}
  {\bibfnamefont {A.~V.}\ \bibnamefont {Zotov}}, \bibinfo {author}
  {\bibfnamefont {A.~A.}\ \bibnamefont {Saranin}}, \bibinfo {author}
  {\bibfnamefont {Y.}~\bibnamefont {Hasegawa}}, \ and\ \bibinfo {author}
  {\bibfnamefont {S.}~\bibnamefont {Hasegawa}},\ }\bibfield  {title} {\enquote
  {\bibinfo {title} {{Unconventional superconductivity in the single-atom-layer
  alloy {S}i(111)-$\sqrt{3}\times\sqrt{3}$-({Tl,Pb})}},}\ }\href {\doibase
  10.1103/PhysRevB.98.134505} {\bibfield  {journal} {\bibinfo  {journal} {Phys.
  Rev. B}\ }\textbf {\bibinfo {volume} {98}},\ \bibinfo {pages} {134505}
  (\bibinfo {year} {2018})}\BibitemShut {NoStop}%
\bibitem [{\citenamefont {Sauls}\ and\ \citenamefont
  {Eschrig}(2009)}]{Sauls_NJP_2009}%
  \BibitemOpen
  \bibfield  {author} {\bibinfo {author} {\bibfnamefont {J~A}\ \bibnamefont
  {Sauls}}\ and\ \bibinfo {author} {\bibfnamefont {M}~\bibnamefont {Eschrig}},\
  }\bibfield  {title} {\enquote {\bibinfo {title} {{Vortices in chiral,
  spin-triplet superconductors and superfluids}},}\ }\href {\doibase
  10.1088/1367-2630/11/7/075008} {\bibfield  {journal} {\bibinfo  {journal}
  {New Journal of Physics}\ }\textbf {\bibinfo {volume} {11}},\ \bibinfo
  {pages} {075008} (\bibinfo {year} {2009})}\BibitemShut {NoStop}%
\bibitem [{\citenamefont {Machida}\ \emph {et~al.}(2018)\citenamefont
  {Machida}, \citenamefont {Kohsaka},\ and\ \citenamefont
  {Hanaguri}}]{Machida_RSI_2018}%
  \BibitemOpen
  \bibfield  {author} {\bibinfo {author} {\bibfnamefont {T.}~\bibnamefont
  {Machida}}, \bibinfo {author} {\bibfnamefont {Y.}~\bibnamefont {Kohsaka}}, \
  and\ \bibinfo {author} {\bibfnamefont {T.}~\bibnamefont {Hanaguri}},\
  }\bibfield  {title} {\enquote {\bibinfo {title} {{A scanning tunneling
  microscope for spectroscopic imaging below 90 mK in magnetic fields up to
  17.5 T}},}\ }\href {\doibase 10.1063/1.5049619} {\bibfield  {journal}
  {\bibinfo  {journal} {Review of Scientific Instruments}\ }\textbf {\bibinfo
  {volume} {89}},\ \bibinfo {pages} {093707} (\bibinfo {year}
  {2018})}\BibitemShut {NoStop}%
\bibitem [{\citenamefont {Caroli}\ \emph {et~al.}(1964)\citenamefont {Caroli},
  \citenamefont {{De Gennes}},\ and\ \citenamefont
  {Matricon}}]{CdGM_PhysLett_1964}%
  \BibitemOpen
  \bibfield  {author} {\bibinfo {author} {\bibfnamefont {C.}~\bibnamefont
  {Caroli}}, \bibinfo {author} {\bibfnamefont {P.~G.}\ \bibnamefont {{De
  Gennes}}}, \ and\ \bibinfo {author} {\bibfnamefont {J.}~\bibnamefont
  {Matricon}},\ }\bibfield  {title} {\enquote {\bibinfo {title} {{Bound Fermion
  states on a vortex line in a type II superconductor}},}\ }\href {\doibase
  https://doi.org/10.1016/0031-9163(64)90375-0} {\bibfield  {journal} {\bibinfo
   {journal} {Physics Letters}\ }\textbf {\bibinfo {volume} {9}},\ \bibinfo
  {pages} {307--309} (\bibinfo {year} {1964})}\BibitemShut {NoStop}%
\bibitem [{\citenamefont {Greco}\ and\ \citenamefont
  {Schnyder}(2018)}]{Greco_PRL_2018}%
  \BibitemOpen
  \bibfield  {author} {\bibinfo {author} {\bibfnamefont {Andr\'es}\
  \bibnamefont {Greco}}\ and\ \bibinfo {author} {\bibfnamefont {Andreas~P.}\
  \bibnamefont {Schnyder}},\ }\bibfield  {title} {\enquote {\bibinfo {title}
  {{Mechanism for Unconventional Superconductivity in the Hole-Doped
  Rashba-Hubbard Model}},}\ }\href {\doibase 10.1103/PhysRevLett.120.177002}
  {\bibfield  {journal} {\bibinfo  {journal} {Phys. Rev. Lett.}\ }\textbf
  {\bibinfo {volume} {120}},\ \bibinfo {pages} {177002} (\bibinfo {year}
  {2018})}\BibitemShut {NoStop}%
\bibitem [{\citenamefont {Ghadimi}\ \emph {et~al.}(2019)\citenamefont
  {Ghadimi}, \citenamefont {Kargarian},\ and\ \citenamefont
  {Jafari}}]{Ghadimi_PRB_2019}%
  \BibitemOpen
  \bibfield  {author} {\bibinfo {author} {\bibfnamefont {Rasoul}\ \bibnamefont
  {Ghadimi}}, \bibinfo {author} {\bibfnamefont {Mehdi}\ \bibnamefont
  {Kargarian}}, \ and\ \bibinfo {author} {\bibfnamefont {S.~A.}\ \bibnamefont
  {Jafari}},\ }\bibfield  {title} {\enquote {\bibinfo {title} {{Competing
  superconducting phases in the interacting two-dimensional electron gas with
  strong Rashba spin-orbit coupling}},}\ }\href {\doibase
  10.1103/PhysRevB.99.115122} {\bibfield  {journal} {\bibinfo  {journal} {Phys.
  Rev. B}\ }\textbf {\bibinfo {volume} {99}},\ \bibinfo {pages} {115122}
  (\bibinfo {year} {2019})}\BibitemShut {NoStop}%
\bibitem [{\citenamefont {Qin}\ \emph {et~al.}(2019)\citenamefont {Qin},
  \citenamefont {Li},\ and\ \citenamefont {Zhang}}]{Qin_NP_2019}%
  \BibitemOpen
  \bibfield  {author} {\bibinfo {author} {\bibfnamefont {Wei}\ \bibnamefont
  {Qin}}, \bibinfo {author} {\bibfnamefont {Leiqiang}\ \bibnamefont {Li}}, \
  and\ \bibinfo {author} {\bibfnamefont {Zhenyu}\ \bibnamefont {Zhang}},\
  }\bibfield  {title} {\enquote {\bibinfo {title} {{Chiral topological
  superconductivity arising from the interplay of geometric phase and electron
  correlation}},}\ }\href {\doibase 10.1038/s41567-019-0517-5} {\bibfield
  {journal} {\bibinfo  {journal} {Nature Physics}\ }\textbf {\bibinfo {volume}
  {15}},\ \bibinfo {pages} {796--802} (\bibinfo {year} {2019})}\BibitemShut
  {NoStop}%
\bibitem [{\citenamefont {Nogaki}\ and\ \citenamefont
  {Yanase}(2020)}]{Nogaki_PRB_2020}%
  \BibitemOpen
  \bibfield  {author} {\bibinfo {author} {\bibfnamefont {Kosuke}\ \bibnamefont
  {Nogaki}}\ and\ \bibinfo {author} {\bibfnamefont {Youichi}\ \bibnamefont
  {Yanase}},\ }\bibfield  {title} {\enquote {\bibinfo {title} {{Strongly
  parity-mixed superconductivity in the Rashba-Hubbard model}},}\ }\href
  {\doibase 10.1103/PhysRevB.102.165114} {\bibfield  {journal} {\bibinfo
  {journal} {Phys. Rev. B}\ }\textbf {\bibinfo {volume} {102}},\ \bibinfo
  {pages} {165114} (\bibinfo {year} {2020})}\BibitemShut {NoStop}%
\bibitem [{\citenamefont {Sun}\ \emph {et~al.}(2021)\citenamefont {Sun},
  \citenamefont {Qin}, \citenamefont {Li},\ and\ \citenamefont
  {Zhang}}]{Sun_PRB_2021}%
  \BibitemOpen
  \bibfield  {author} {\bibinfo {author} {\bibfnamefont {Shuwen}\ \bibnamefont
  {Sun}}, \bibinfo {author} {\bibfnamefont {Wei}\ \bibnamefont {Qin}}, \bibinfo
  {author} {\bibfnamefont {Leiqiang}\ \bibnamefont {Li}}, \ and\ \bibinfo
  {author} {\bibfnamefont {Zhenyu}\ \bibnamefont {Zhang}},\ }\bibfield  {title}
  {\enquote {\bibinfo {title} {{Chiral topological superconducting state with
  Chern number $\mathcal{C}$ $=$ $\ensuremath{-}2$ in
  ${\mathrm{Pb}}_{3}\mathrm{Bi}/\mathrm{Ge}(111)$}},}\ }\href {\doibase
  10.1103/PhysRevB.103.235149} {\bibfield  {journal} {\bibinfo  {journal}
  {Phys. Rev. B}\ }\textbf {\bibinfo {volume} {103}},\ \bibinfo {pages}
  {235149} (\bibinfo {year} {2021})}\BibitemShut {NoStop}%
\bibitem [{\citenamefont {Kohsaka}\ \emph {et~al.}(2017)\citenamefont
  {Kohsaka}, \citenamefont {Machida}, \citenamefont {Iwaya}, \citenamefont
  {Kanou}, \citenamefont {Hanaguri},\ and\ \citenamefont
  {Sasagawa}}]{Kohsaka_PRB_2017}%
  \BibitemOpen
  \bibfield  {author} {\bibinfo {author} {\bibfnamefont {Y.}~\bibnamefont
  {Kohsaka}}, \bibinfo {author} {\bibfnamefont {T.}~\bibnamefont {Machida}},
  \bibinfo {author} {\bibfnamefont {K.}~\bibnamefont {Iwaya}}, \bibinfo
  {author} {\bibfnamefont {M.}~\bibnamefont {Kanou}}, \bibinfo {author}
  {\bibfnamefont {T.}~\bibnamefont {Hanaguri}}, \ and\ \bibinfo {author}
  {\bibfnamefont {T.}~\bibnamefont {Sasagawa}},\ }\bibfield  {title} {\enquote
  {\bibinfo {title} {{Spin-orbit scattering visualized in quasiparticle
  interference}},}\ }\href {\doibase 10.1103/PhysRevB.95.115307} {\bibfield
  {journal} {\bibinfo  {journal} {Phys. Rev. B}\ }\textbf {\bibinfo {volume}
  {95}},\ \bibinfo {pages} {115307} (\bibinfo {year} {2017})}\BibitemShut
  {NoStop}%
\bibitem [{\citenamefont {Kohsaka}\ \emph {et~al.}(2007)\citenamefont
  {Kohsaka}, \citenamefont {Taylor}, \citenamefont {Fujita}, \citenamefont
  {Schmidt}, \citenamefont {Lupien}, \citenamefont {Hanaguri}, \citenamefont
  {Azuma}, \citenamefont {Takano}, \citenamefont {Eisaki}, \citenamefont
  {Takagi}, \citenamefont {Uchida},\ and\ \citenamefont
  {Davis}}]{Kohsaka_Science_2007}%
  \BibitemOpen
  \bibfield  {author} {\bibinfo {author} {\bibfnamefont {Y.}~\bibnamefont
  {Kohsaka}}, \bibinfo {author} {\bibfnamefont {C.}~\bibnamefont {Taylor}},
  \bibinfo {author} {\bibfnamefont {K.}~\bibnamefont {Fujita}}, \bibinfo
  {author} {\bibfnamefont {A.}~\bibnamefont {Schmidt}}, \bibinfo {author}
  {\bibfnamefont {C.}~\bibnamefont {Lupien}}, \bibinfo {author} {\bibfnamefont
  {T.}~\bibnamefont {Hanaguri}}, \bibinfo {author} {\bibfnamefont
  {M.}~\bibnamefont {Azuma}}, \bibinfo {author} {\bibfnamefont
  {M.}~\bibnamefont {Takano}}, \bibinfo {author} {\bibfnamefont
  {H.}~\bibnamefont {Eisaki}}, \bibinfo {author} {\bibfnamefont
  {H.}~\bibnamefont {Takagi}}, \bibinfo {author} {\bibfnamefont
  {S.}~\bibnamefont {Uchida}}, \ and\ \bibinfo {author} {\bibfnamefont {J.~C.}\
  \bibnamefont {Davis}},\ }\bibfield  {title} {\enquote {\bibinfo {title} {{An
  Intrinsic Bond-Centered Electronic Glass with Unidirectional Domains in
  Underdoped Cuprates}},}\ }\href {\doibase 10.1126/science.1138584} {\bibfield
   {journal} {\bibinfo  {journal} {Science}\ }\textbf {\bibinfo {volume}
  {315}},\ \bibinfo {pages} {1380--1385} (\bibinfo {year} {2007})}\BibitemShut
  {NoStop}%
\bibitem [{\citenamefont {Kohsaka}(2021)}]{Kohsaka_RSI_2021}%
  \BibitemOpen
  \bibfield  {author} {\bibinfo {author} {\bibfnamefont {Yuhki}\ \bibnamefont
  {Kohsaka}},\ }\bibfield  {title} {\enquote {\bibinfo {title} {{Removing
  background and estimating a unit height of atomic steps from a scanning probe
  microscopy image using a statistical model}},}\ }\href {\doibase
  10.1063/5.0038852} {\bibfield  {journal} {\bibinfo  {journal} {Rev. Sci.
  Instrum.}\ }\textbf {\bibinfo {volume} {92}},\ \bibinfo {pages} {033702}
  (\bibinfo {year} {2021})}\BibitemShut {NoStop}%
\end{thebibliography}%

\clearpage
\begin{figure*}[t]
    \begin{center}
    \includegraphics[width=17cm]{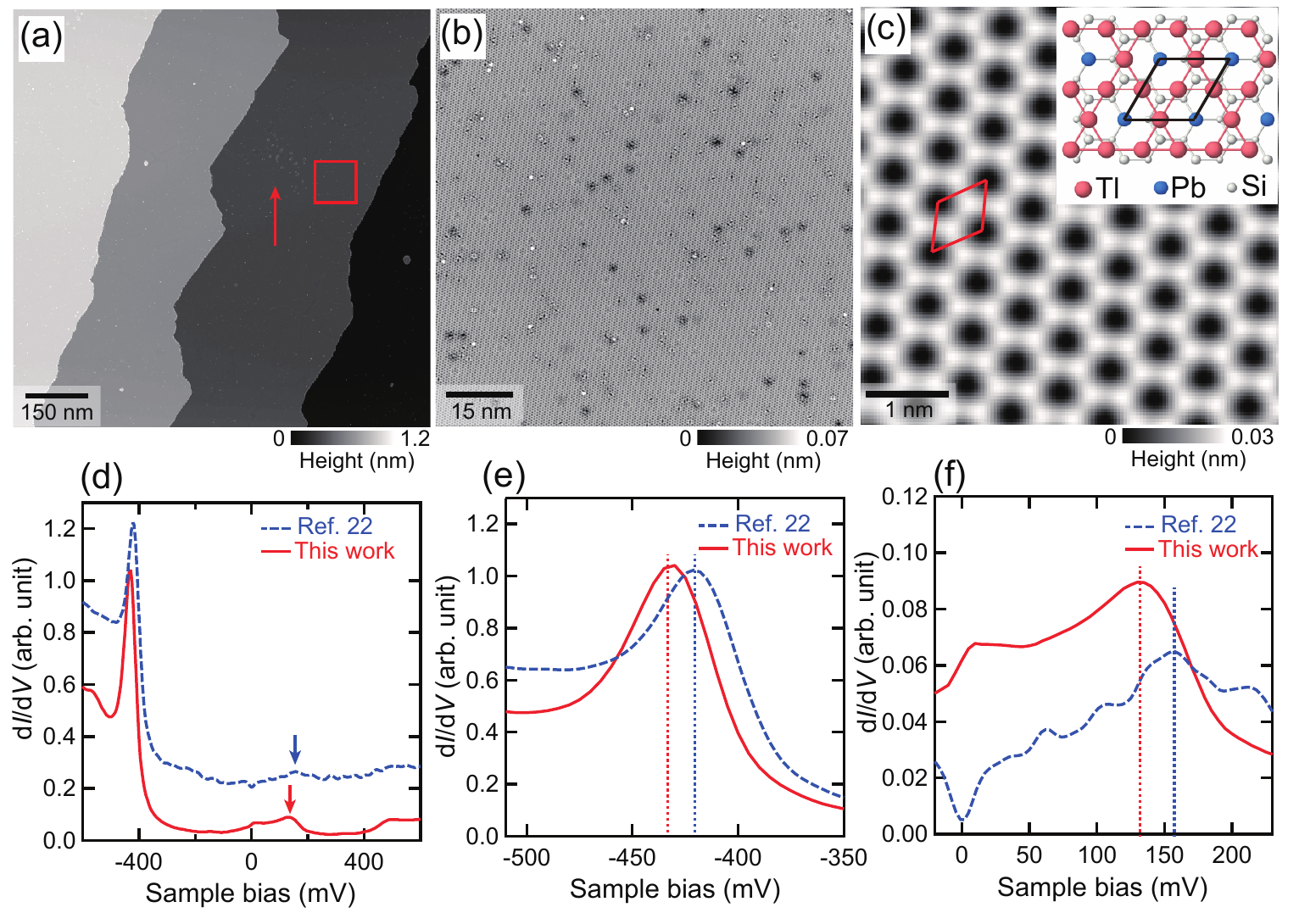}
    \end{center}
    \caption{    
    (a) An STM topographic image over a $1~\mu\rm{m}~\times~1~\mu\rm{m}$ field of view.
    The image background including the tilt of sample and the creep of scanning piezo actuators is subtracted using the statistical analysis of topographic image developed in Ref. \onlinecite{Kohsaka_RSI_2021}.
The feedback loop was stabilized at the set-point voltage $V_{\rm{s}}~=~1~\rm{V}$ and current $I_{\rm{s}}~=~10~\rm{pA}$. 
The arrow indicates the path along which the series of superconducting-gap spectra shown in Fig. 2(b) was taken. 
    (b) A magnified topographic image over the $94~\rm{nm}~\times~94~\rm{nm}$ field of view marked by the box in (a), ($V_{\rm{s}}~=~300~\rm{mV}$ and $I_{\rm{s}}~=~1~\rm{nA}$).
    (c) Atomic resolution topographic image over a $5~\rm{nm}~\times~5~\rm{nm}$ field of view ($V_{\rm{s}}~=~300~\rm{mV}$ and $I_{\rm{s}}~=~1~\rm{nA}$). 
    A rhombus represents the unit cell.
    Inset: A schematic illustration of the crystal structure of the Si(111)-$\sqrt{3}\times\sqrt{3}$-(Tl,Pb).
    Magenta, cyan, and gray spheres indicate Tl, Pb, and Si atoms, respectively. A black rhombus represents the unit cell.
    (d) Red curve denotes the tunneling spectrum taken at 4.4~K.
    The spectrum of the sample used in Ref.~\onlinecite{Nakamura_PRB_2018} is shown in blue for comparison.
    A vertical offset by 0.2 is added for clarity. 
    Arrows indicate the hump structures around +150~mV.
    (e) Magnified spectra near the peak at $\sim -400$~mV without the vertical shift.
    (f) Magnified spectra near the hump at $\sim +150$~mV without the vertical shift.
    In (e) and (f), red and blue dashed vertical lines indicate the locations of peaks and humps.
    }
\end{figure*}

\clearpage
\begin{figure*}[t]
    \begin{center}
    \includegraphics[width=17cm]{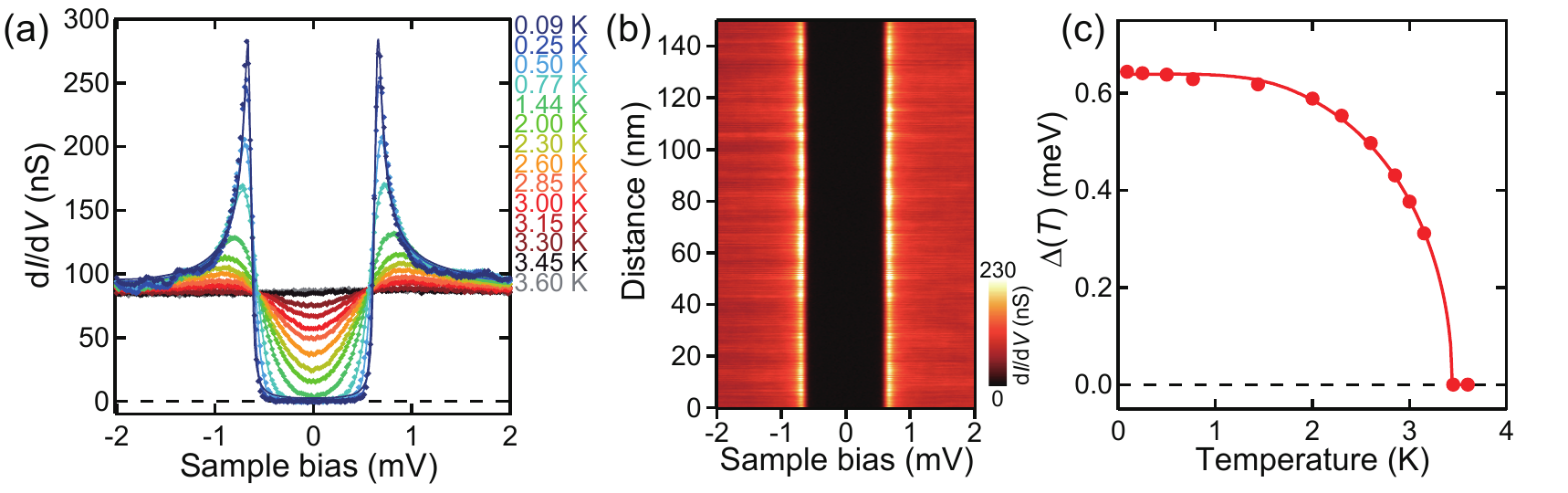}
    \end{center}
    \caption{
    (a) Temperature dependence of the tunneling spectra taken at the set point of $V_{\rm{s}}~=~10$~mV, $I_{\rm{s}}~=~1$~nA and with the lock-in modulation of $V_{\rm{mod}}~=~14.14~\mu$V$_{\rm{rms}}$.
    The colors of the spectra correspond to those of the measured temperatures indicated on the right side of the graph. 
    Dots and lines represent the raw data points and fitting results by Dynes function.
    (b) Spatial variation of the tunneling spectra at 90~mK along the red arrow in Fig. 1(a). $V_{\rm{s}}$, $I_{\rm{s}}$ and $V_{\rm{mod}}$ are the same in (a).
    (c) Temperature dependence of the superconducting gap size extracted by fitting the spectra in (a).
    }
\end{figure*}

\clearpage
\begin{figure*}[t]
    \begin{center}
    \includegraphics[width=17cm]{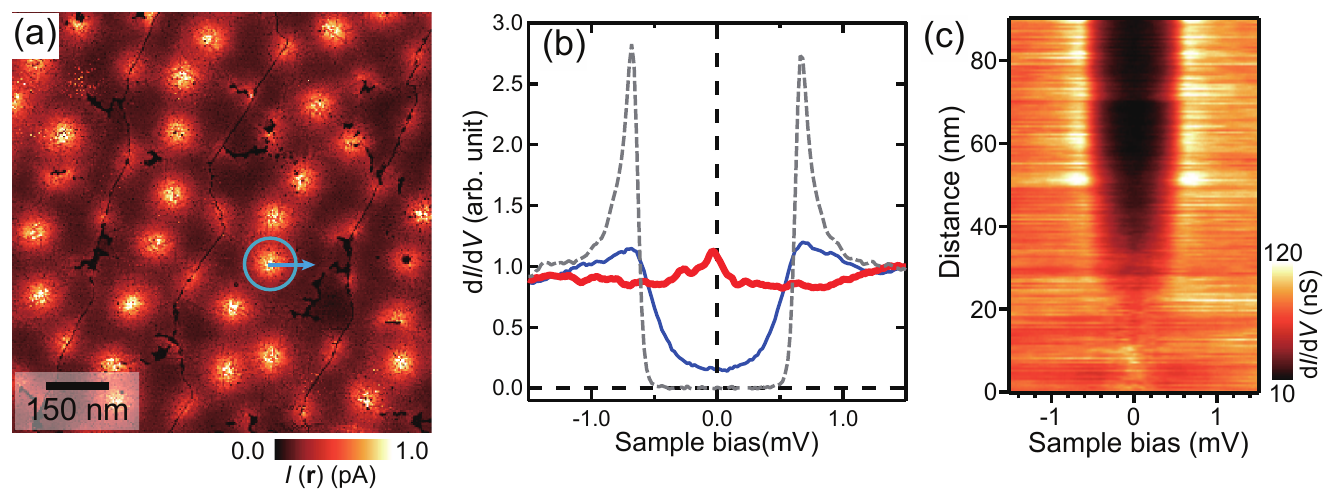}
    \end{center}
    \caption{    
    (a) A current image $I(\bm{r}, E~=~300~\mu\rm{eV})$ taken at the set point of $V_{\rm{s}}~=~500$~mV, $I_{\rm{s}}~=~1$~nA on the same field of view as Fig. 1(a). 
    (b) Tunneling spectra at the vortex core marked by a circle in (a). 
    Gray dashed, red, and blue solid lines indicate the spectrum taken under 0~T, at the vortex core and 80 nm away from the core under 0.1~T, respectively.
    The spectra are normalized by the conductance at +1.5 mV. 
    (c) Line profile of the tunneling spectra along a path shown by an arrow in (a). 
    All the data are taken at 90~mK.
    }
\end{figure*}

\clearpage
\begin{figure*}[t]
    \begin{center}
    \includegraphics[width=17cm]{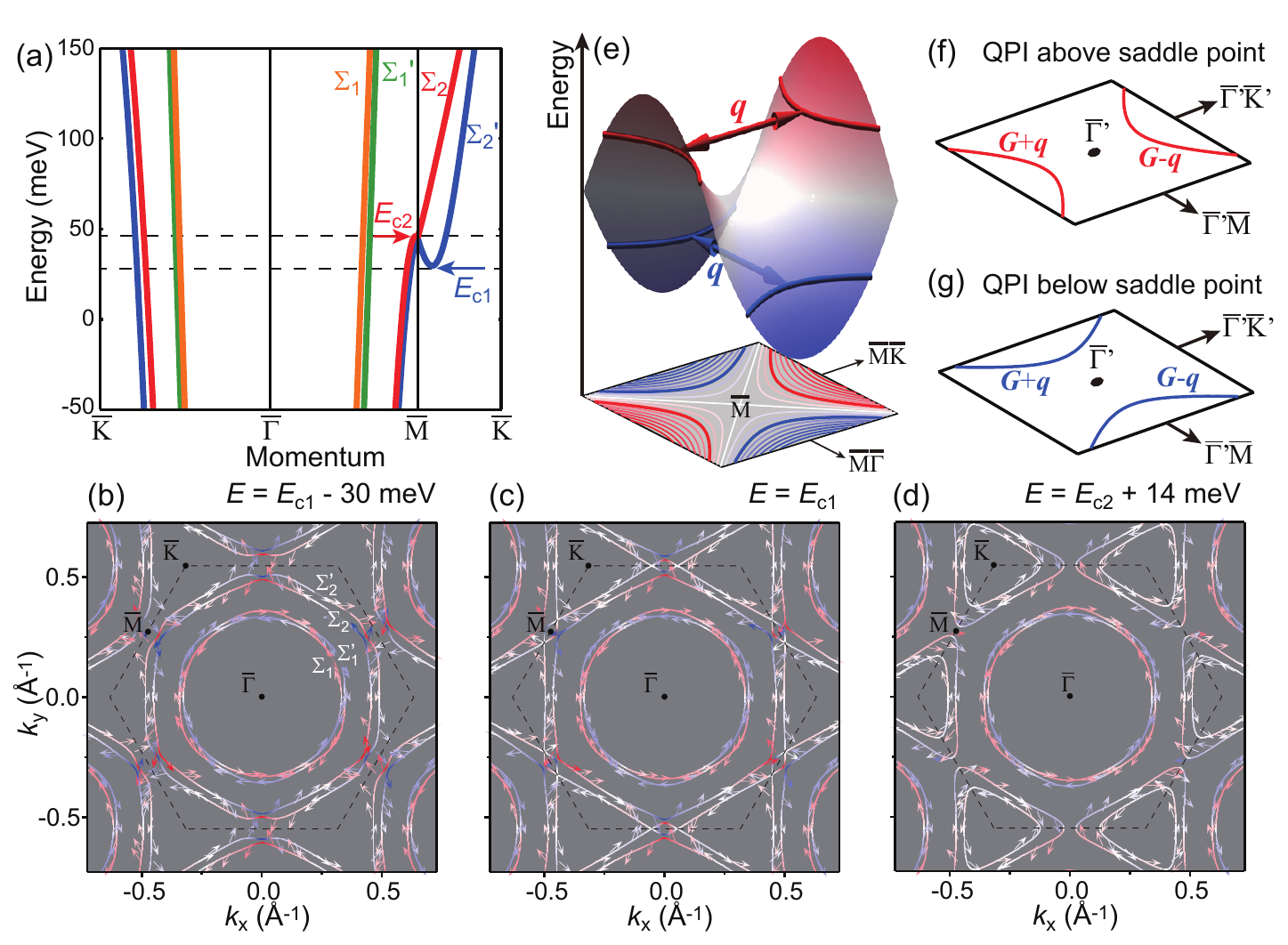}
    \end{center}
    \caption{   
     (a) A band structure near the Fermi level obtained by the tight-binding model described in the Appendix B. Blue and red arrows represent the energies at the saddle point ($E_{\rm{c1}}$) and the band edge at $\overline{\rm{M}}$ point ($E_{\rm{c2}}$), respectively.
     As described in the main text, the actual Fermi levels of the present sample and previous (Ref. \onlinecite{Nakamura_PRB_2018}) samples lie around +30 meV (near $E_{\rm{c1}}$) and +15 meV, respectively.
     (b)-(d) Constant energy contours at representative energies. Arrows represent the direction of the in-plane components of the spin expectation values. The colors of the arrows show the out-of-plane spin component.
     Dashed lines denote the first Brillouin zone (BZ).
     (e) Schematic band structure near the saddle point. Red and blue arrows represent representative QPI scattering vectors above and below the saddle point, respectively. Bottom panel indicates the constant energy contours. 
     (f) and (g) Expected QPI patterns near the Bragg point above and below the saddle point, respectively.
    }
\end{figure*}

\clearpage
\begin{figure*}[t]
    \begin{center}
    \includegraphics[width=17cm]{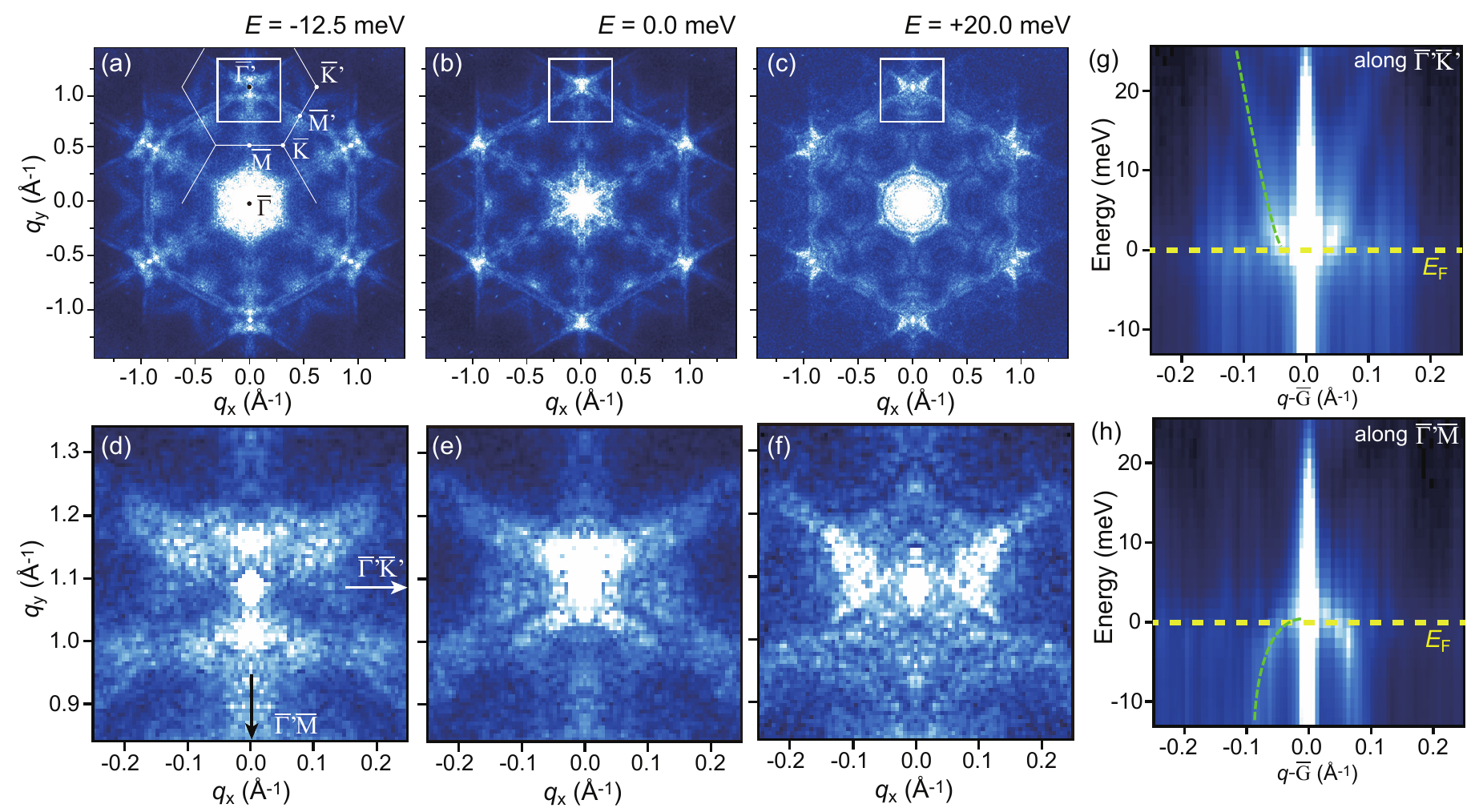}
    \end{center}
    \caption{    
    (a)-(c) QPI patterns at representative energies appearing in the symmetrized Fourier-transformed conductance maps. The data was taken at 4.4~K with $V_{\rm{s}}$~=~50~mV, $I_{\rm{s}}$~=~1.5~nA and $V_{\rm{mod}}$~=~0.88~mV$_{\rm{rms}}$.
    (d)-(f) The magnified QPI patterns around the $\overline{\rm{\Gamma'}}$ point marked by white boxes in (a)-(c), respectively.
    (g) and (h) Line profiles of the QPI patterns around the Bragg point along the $\overline{\rm{\Gamma'}}\overline{\rm{M}}$ and $\overline{\rm{\Gamma'}}\overline{\rm{K'}}$directions, respectively. In (g) and (h) dashed green lines are eye guides of the observed branches.
    }
\end{figure*}

\clearpage
\begin{figure*}[t]
    \begin{center}
    \includegraphics[width=17cm]{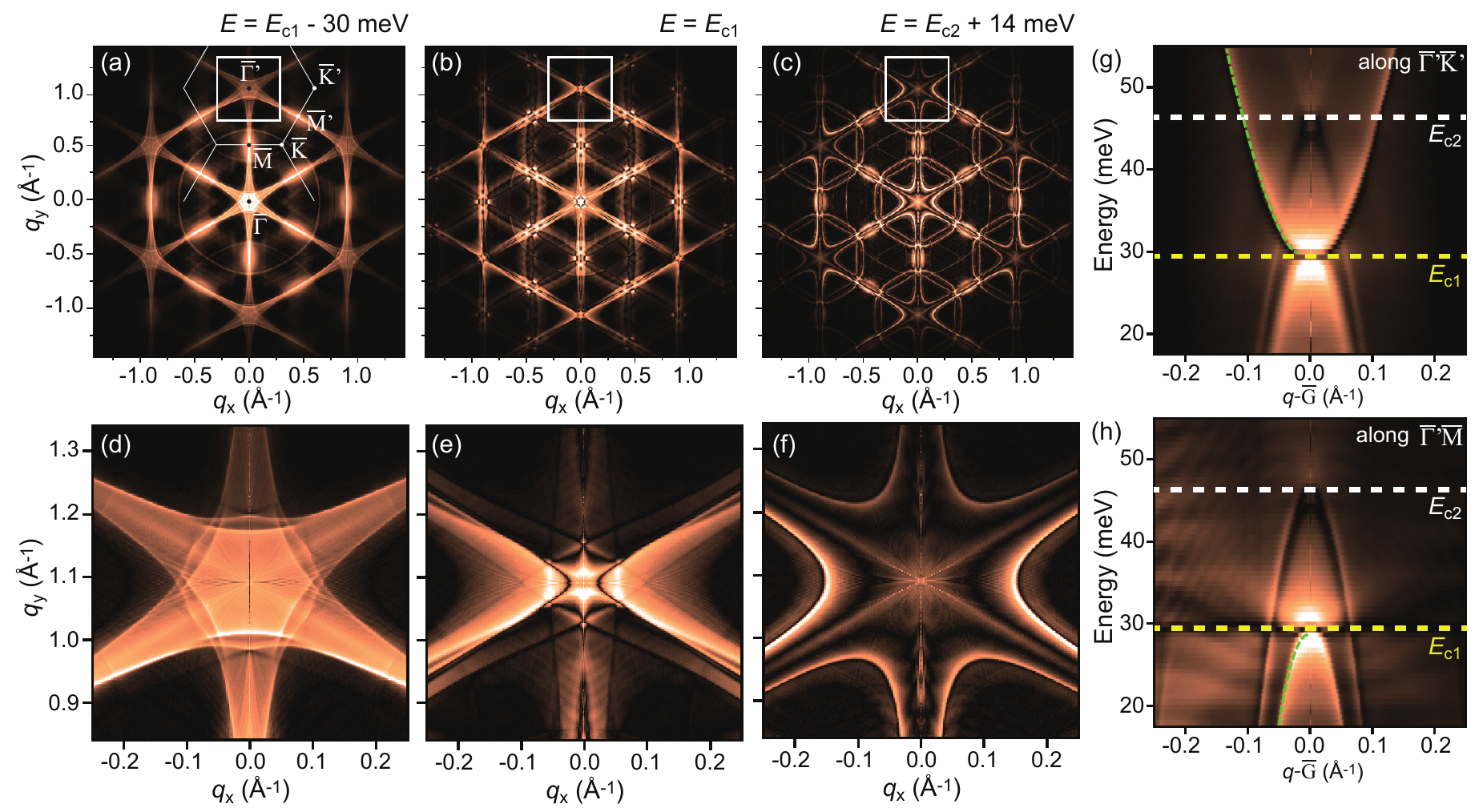}
    \end{center}
    \caption{  
    (a)-(c) Simulated QPI patterns at representative energies. 
    (d)-(f) The magnified images around the $\overline{\rm{\Gamma'}}$ point marked by white boxes in (a)-(c), respectively.
    (g) and (h) Line profiles of the simulated QPI patterns around the $\overline{\rm{\Gamma'}}$ point along the $\overline{\rm{\Gamma'}}\overline{\rm{M}}$ and $\overline{\rm{\Gamma'}}\overline{\rm{K'}}$directions, respectively.  
    In (g) and (h) dashed green lines are eye guides of the branches corresponding to the observed ones.
        }
\end{figure*}

\clearpage
\begin{figure}[t]
    \begin{center}
    \includegraphics[width=8cm]{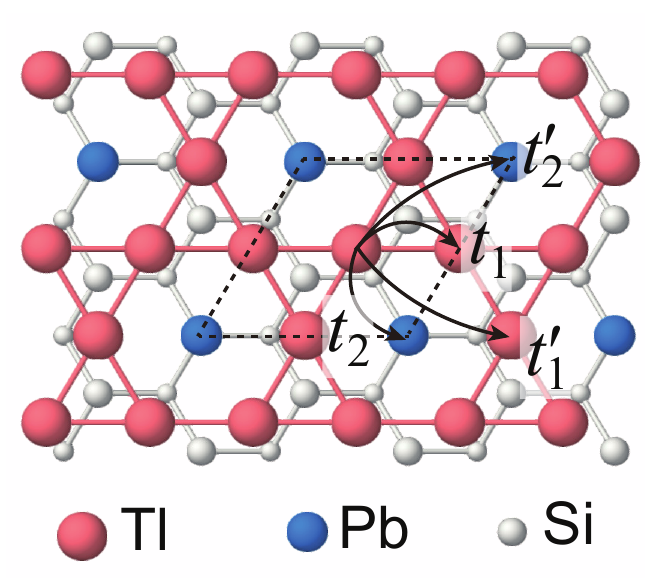}
    \end{center}
    \caption{  
    Sketch of hopping parameters used in the tight binding model.}
\end{figure}

\end{document}